\documentclass[iop,twocolumn,twocolappendix,numberedappendix]{aastex63}%{emulateapj}
\usepackage{epsfig}
\usepackage{amsmath}
\usepackage{amssymb}
\usepackage{natbib}
\usepackage{hyperref}
\usepackage{graphicx}
\usepackage{mathrsfs}
\usepackage{times}
  \usepackage[flushleft]{threeparttable}

\newcommand{\vect}[1]{\boldsymbol{#1}}

\begin{document}
\defcitealias{GLB}{GLB}
\title{Molecular clouds as gravitational instabilities in rotating disks: a modified stability criterion
} 
\author{Sharon E. Meidt}
\affiliation{Sterrenkundig Observatorium, Universiteit Gent, Krijgslaan 281 S9, B-9000 Gent, Belgium}
\begin{abstract}
Molecular gas disks are generally Toomre stable ($Q_T>$1) and yet clearly gravitationally unstable to structure formation as evidenced by the existence of molecular clouds and ongoing star formation. This paper adopts a 3D perspective to obtain a general picture of instabilities in flattened rotating disks, using the 3D dispersion relation to describe how disks evolve when perturbed over their vertical extents. By explicitly adding a vertical perturbation to an unperturbed equilibrium disk, stability is shown to vary with height above the mid-plane. Near to $z$=0 where the equilibrium density is roughly constant, instability takes on a Jeans-like quality, occurring on scales larger than the Jeans length and subject to a threshold $Q_M=\kappa^2/(4\pi G\rho)=1$ or roughly $Q_T\approx 2$. Far from the mid-plane, on the other hand, stability is pervasive, and the threshold for the total disk (out to $z=\pm\infty$) to be stabilized is lowered to $Q_T=1$ as a consequence. In this new framework, gas disks are able to fragment through partial 3D instability even where total 2D instability is suppressed. The growth rates of the fragments formed via 3D instability are comparable to, or faster than, Toomre instabilities. The rich structure in molecular disks on the scale of 10s of pc can thus be viewed as a natural consequence of their 3D nature and their exposure to a variety of vertical perturbations acting on roughly a disk scale height, i.e. due to their situation within the more extended galaxy potential, participation in the disk-halo flow, and exposure to star formation feedback. \\
\end{abstract}

%\maketitle
\section{Introduction\label{sec:intro}}
\setcounter{footnote}{0}
One of the long-standing idiosyncracies of star formation theory is that the molecular gas disks of galaxies where stars are formed -- and which are rich in multi-scale structure -- lie largely above the Toomre `Q' threshold \citep{toomre} used to predict instability and fragmentation in rotating disks \citep[see][]{leroy08,romeowiegert,romeomogotsi, elm11}.  Yet there is no denying the appeal of a picture for star formation in which the initial stages involve passing the threshold for gravitational instability after which feedback from subsequent star formation returns the molecular medium to the brink of gravitational instability \citep[so-called `Q' regulation; e.g][]{silk97, ko01, hopkins}.  

The modern observation that molecular disks have $Q$$>$1 \citep[i.e.][]{kenn89, MK01, leroy08} has thus prompted a number of revisions of the criterion \citep[e.g.][]{romeo10,elm11,romeowiegert,grivgedalin, romeoagertz, agertz15}.  These follow along the lines of studies that take into account magnetic fields \citep[e.g.][]{chandrasekhar54,balbus,elm87,elm94,gammie96,ko01}, cooling and the dissipative nature of turbulent gas \citep{elm89, gammie01}, and the role of non-axisymmetry on gas disk stability \citep{GLBb,JT66}.  These all suggest, either phenomenologically, analytically or numerically, that instability and fragmentation may be possible above the $Q$=1 threshold.  

In another school of thought, the Toomre instability in its traditional form is only indirectly relevant to the star formation process \citep{koda,elm11}, i.e. because the instability is really taking place in an inherently multi-component galaxy disk \citep{jogsolomon,bertinromeo,wangsilk94,ko07,romeowiegert} and the product of the instability under typical conditions in star-forming disks is large-scale structures that organize the molecular medium, rather than the molecular clouds \citep{elm11}.  In this school, molecular cloud formation requires alternative channels \citep[see review by][and references therein]{dobbs14} and these must necessarily act on short timescales \citep{maclow17}, given the developing consensus from both simulations and observations that molecular clouds are rapidly destroyed by early stellar feedback together with galactic shear (see review by \citealt{dale15} and e.g. \citealt{kk018,meidt15,chevance20,chevance21,kim21}).   

From another perspective, it may not be surprising that the Toomre criterion fails to be predictive of molecular structures that are 10s of parsecs in scale -- close to the disk scale height -- given that it was designed to describe stability to large-scale perturbations confined specifically to thin disks.  Indeed, the perturbations that are typically considered are most often confined to two-dimensions, approximating the disk's internal response (mediated by self-gravity) to some local impulse, again acting from within.  These are the types of perturbations relevant for describing the stability of density wave perturbations \citep{linshu,toomre}.  

In contrast to collisionless rotating stellar disks with smooth vertical profiles, however, molecular gas disks reveal their susceptibility to impulses and perturbations that are not necessarily 2D, restricted to the mid-plane, or tied to the disk's vertical extent, e.g. triggered by phase transitions, star formation feedback and the disk-halo flow \citep[e.g.][]{fraternalibinney06,walch2015,elm14}, or related to non-axisymmetric structures in the surrounding stellar disk or interaction with the local environment \citep[e.g. ram pressure stripping; ][]{vollmer,lee17}.

In this paper alternative vertical perturbations are proposed, designed with (molecular) gas disks embedded in thicker gas and stellar disks in mind, and used to derive an analytical condition for disk stability on scales near the disk scale height.  
The paper centers on the 3D dispersion relation, which relates the evolution of the perturbation to the vertical and radial motions that develop from self-gravity, rotation and gas pressure.  Readers primarily interested in the application of the new framework to the stability of molecular disks are pointed to $\S$~\ref{sec:3Dv2D}.  The interested reader can find the details of the derivation of the 3D and 2D dispersion relations in $\S\S$~\ref{sec:framework} and \ref{sec:2dstability}.  A summary of how disks are shown to behave in the presence of different types of perturbations (examined in detail in $\S\S$~\ref{sec:framework} and ~\ref{sec:2dstability}) is given at the end of $\S$~\ref{sec:2dstability}.  There the reader can also obtain an overview of the two main modes of instability in 3D flattened rotating disks: 2D Toomre instability and the 3D instability endemic to the mid-plane identified in this work.  

In more detail, after introducing the framework used to obtain solutions to the 3D linearized equations of motion in $\S\S$~\ref{sec:framework21} and \ref{sec:equationsofmotion}, the 3D dispersion relation is obtained and used to assess stability near to and far from the disk mid-plane in $\S$~\ref{sec:3ddispersionRelation}.  %\ref{sec:verticalonly}.  
Then in $\S$~\ref{sec:2dstability}, following \cite{toomre} and \cite{GLB} the 2D version of the dispersion relation is used to determine the conditions for instability (perturbation growth) in a number of scenarios.  The threshold calculated by GLB in the case of infinite vertical perturbations with no phase variation is recovered in $\S$~\ref{sec:infGLB}.  The impact of wave-like behavior on this threshold is considered in $\S$~\ref{sec:infwave}.  Then in $\S$~\ref{sec:finWKB} the Toomre criterion is obtained using a vertical perturbation that is both wave-like, with a specific relation between the vertical and radial wavenumbers, and also extended relative to the disk scale height $h$.  Finally, a modified, higher threshold is obtained for wave and non-wave perturbations near the mid-plane. To assess the prominence of fragments formed via gravitational instability in these different scenarios, in $\S$~\ref{sec:3Dv2D}
%\ref{sec:growthrates} 
the growth rates of unstable 3D perturbations are calculated over a range of spatial scales.  
 
\section{Three-dimensional instability in rotating disks}\label{sec:framework}
\subsection{The Basic Framework}\label{sec:framework21}
To examine the conditions that lead to gravitational instability in 3D rotating gas disks we adopt the idealized configuration proposed by \cite[][hereafter GLB]{GLB}, in which the disk is infinitely extended in the radial and vertical directions but significantly compressed in the vertical direction parallel to the axis of rotation.  The gas in this disk is assumed to be approximately isothermal and undergo non-uniform rotation at a rate $\Omega$ that depends on galactocentric radius.    

With this framework, 
we obtain the dispersion relation for density perturbations propagating in the gas disk by combining the continuity equation

 \begin{equation}
\frac{\partial\rho}{\partial t}=\vect{\nabla}\cdot(\rho\vect{v})=0
\end{equation}
with solutions to the Euler equations of motion for the rotating disk plus a small perturbation, 

\begin{equation}
\frac{\partial\vect{v}}{\partial t}+(\vect{v}\cdot\vect{\nabla})\vect{v} = -\frac{1}{\rho}\vect\nabla p-\vect{\nabla}\Phi. \label{eq:EOM}
\end{equation}
Here $\rho$ is the gas density, $p$ is the thermal plus turbulent gas pressure \citep[following][]{chandrasekhar51} and the gravitational potential $\Phi$ represents gas self-gravity together with a possible background potential defined by a surrounding distribution of gas, stars and dark matter.  

Although it has become common to only consider the linearized equations of motion in 2D polar coordinates, adopting perturbations that involve no motion in the vertical direction, a full 3D treatment in cylindrical coordinates has also been previously considered \citep[i.e.][]{GLB}.  
We follow the latter approach, and employ a number of the techniques common to 2D and 3D calculations.  For one, the equations of motion are typically satisfied by adopting an $m$-mode perturbation of the form $\propto$ $\exp{i (m\phi-\omega t+\vect{k}\cdot\vect{r})}$ propagating in the direction $\vect{r}$ with wavenumber $\vect{k}$ where $\omega$ is the oscillation frequency of the mode \citep[e.g.][]{toomre,linshu,BT}.  
The unstable growing modes can then be identified by the condition $\omega^2$$<$0.  

The equations of motion are also typically simplified using the WKB (Wentzel-Kramers-Brillouin) approximation, in which the phase of the radial perturbation is assumed to be rapidly varying ($k R$$>>$1) so that the variation in the perturbation amplitude is negligible and terms of order $1/R$ are neglected in favor of those of order $k$.  

The two distinguishing features of this work (described in more detail below) are a non-zero vertical velocity dispersion and the possibility of motion in the vertical direction described by vertical perturbations that explicitly include phase variation (wave-like behavior) and which are either infinite in extent or finite and described in the WKB approximation.  Thus, the approach is most similar to \citetalias{GLB}, except that here a broader set of perturbations are considered and stability is examined from both the 2D and 3D perspectives.  That is, before obtaining the 2D dispersion relation, this work uses the 3D dispersion relation 
in a number of different regimes to make transparent predictions for the scales of instabilities and assess how instability various with height above the mid-plane.  Then, following \cite{toomre} and  \citetalias{GLB}, the 2D version of the dispersion relation is obtained to identify the conditions for the overall stability of the disk.

\subsubsection{Vertical Motions in the Unperturbed Disk}
A major motivation for including the vertical dimension is to obtain a realistic description of molecular gas disks in which turbulent gas motions are three-dimensional and nearly isotropic and in which the embedded clouds are triaxial.  
As will be shown in $\S$~\ref{sec:largeh1}, the 3D dispersion relation derived in this work approaches the 2D Lin-Shu dispersion relation in the limit $\sigma_z\rightarrow 0$.  
For the more general scenario of interest here, both vertical and radial components of motion are allowed, each given  the following 1D velocity dispersion \citep[i.e.][]{chandrasekhar51}

\begin{equation}
\sigma_{\rm i}^2=v_s^2+\sigma_{\rm turb,i}^2
\end{equation}
which combines the sound speed in the gas $v_s$ with turbulent motions $\sigma_{\rm turb,i}$ in direction $i$.  Although we allow that $\sigma_z$$\neq$$\sigma_r$, velocity dispersions in the gas are generally assumed to be isotropic.  

The turbulent motions in the gas are envisioned as arising from two main sources that combine to yield an effective (non-thermal) pressure that places the disk in dynamical equilibrium.  Star formation feedback (plus turbulent dissipation) is assumed to set a base pressure $p_{\rm FB}$.  This combines with the effective pressure $p_{\rm eff}$ set up by the averaged kinematic response of many individual fluid elements to the force set up by the remainder between the gravitational force and the gradient in the baseline $p_{\rm FB}$.  For a given $p_{\rm FB}$, $p_{\rm eff}$ is thus the pressure that maintains the gas disk in overall equilibrium.  Thus, in what follows, dynamical equilibrium is applied even when feedback-driven turbulent pressure is either zero or very small due to the absence star formation.  

For this equilibrium scenario, unless otherwise noted, the unperturbed vertical density distribution is envisioned as falling between the self-gravitating profile

\begin{equation}
\rho_0(z)=\rho_c sech^2(z/h),\label{eq:vertdistrib1}
\end{equation}

where $h=\sigma_z/(2\pi G\rho_c)^{1/2}$, and 
\begin{equation}
\rho_0(z)=\rho_c e^{-z^2/h^2}\label{eq:vertdistrib2}
\end{equation}
in the presence of a dominant external potential generated by the background distribution with density $\rho_b$, where $h=\sigma_z/(4\pi G\rho_b)^{1/2}$.  The equilibrium vertical velocity dispersion associated with these profiles is constant (independent of $z$).  

\subsubsection{Vertical and Radial Perturbations}\label{sec:perturbations}
Allowing the gas disk to have some thickness, we now wish to incorporate realistic perturbations that are compatible with the sorts of influences that a gas disk experiences.   The goal with these perturbations is not to represent a specific process but rather to invoke a generic influence that is non-negligible away from the galactic mid-plane in a manner that satisfies the linearized equations of motion in 3D.   
Thus, we adopt wave perturbations of the form 

\begin{equation}
\Phi_1(R,\phi,z,t)=Re[\mathcal{F}(R,z)e^{i(m\phi-\omega t)}e^{ik R}e^{i k_z z}]  \label{eq:pertex}
\end{equation}
or
\begin{equation}
\Phi_1(R,\phi,z,t)=i Im[\mathcal{F}(R,z)e^{i(m\phi-\omega t)}e^{ik R}e^{i k_z z}]  \label{eq:pertex2}
\end{equation}
where the wavenumbers $k$=$2\pi/\lambda_r$ and $k_z$=$2\pi/\lambda_z$ describe the wavelengths of the perturbation in the radial and vertical directions, respectively.  

As is typical, these perturbations are assumed to satisfy the WKB approximation in the radial direction, i.e. $\partial \mathcal{F}(R,z)/\partial R= \mathcal{F}(R,z)/R_{p}<<ik\mathcal{F}(R,z)$ where $R_{p}$ is the characteristic scale over which the amplitude of the perturbation varies in the radial direction.  In practice this is adopted as the criterion $kR>>1$. 

In the vertical direction, three different scenarios are considered.  In the first two scenarios, the perturbations are assumed to be infinite, pervading every location in the disk, but either the wave nature is neglected in the vertical direction and so $k_z=0$ (the first scenario, as considered by GLB) or $k_z$ is assumed to be a constant and independent of height $z$ above the mid-plane (the second scenario).  In both of these cases, the amplitude of the perturbation must vary faster in the vertical direction than the vertical density variation in the unperturbed disk, in order that the perturbation remains small with respect to $\rho_0$ everywhere ($\rho_1<<\rho_0$).   Defining $Z_d=(d\textrm{ln}\Phi_0/dz)^{-1}$ and $Z_p$=$(d\textrm{ln} \mathcal{F}(R,z)/dz)^{-1}$, then for these infinite perturbations, $Z_p\lesssim Z_d$.  As in $\S\S$ \ref{sec:infGLB} and \ref{sec:infwave} this can be cast in terms of the vertical gradients in the unperturbed and perturbed densities, i.e. $z_p=z_d$ with $z_d=(d\textrm{ln}\rho_0/dz)^{-1}$ and $z_p$=$(d\textrm{ln} \rho_1/dz)^{-1}$. 

The wave type of these infinite perturbations are not examined in the WKB approximation, given that they entail rapid variation in the perturbed amplitude.  Thus, any infinitely extended 3D perturbation considered in this work satisfies the most general form of the perturbed Poisson's equation 

\begin{equation}
\Phi_1=\frac{4\pi G\rho_1}{-k_{\textrm{plane}}^2-k_z^2+T^2}
\end{equation}
where $k_{\textrm{plane}}^2\equiv k^2+m^2/R^2$ and

\begin{eqnarray}
T^2&\equiv&(\nabla_z^2\Phi_1)/\Phi_1+k_z^2\nonumber\\
&=&\nabla_z\left(\frac{1}{Z_p}\right)+\frac{1}{Z_p^2}+\frac{2ik_z}{Z_p}.
\end{eqnarray}

In the third scenario, wave perturbations are restricted to a finite extent above and below the mid-plane.  Such perturbations can be studied without the requirement that $Z_p=Z_d$ (or $z_p=z_d$), since they are not at risk of becoming non-negligible as the unperturbed disk density drops towards $z\rightarrow\pm\infty$.  These perturbations could vary arbitrarily in amplitude in the  $z$ direction (as long as this variation is negligible with respect to the perturbation's phase variation, in the WKB approximation) and would thus not need to satisfy $k_z z_d>>1$, or necessarily share the overall variation of the unperturbed disk.  The amplitude of the perturbation might instead vary much more slowly than $\rho_0$, perhaps tied to the density distribution of an embedding disk or a process active therein, for instance.  In this scenario, the WKB approximation is invoked as $k_z z_p>>1$ (or $T\approx 0$).\footnote{These finite perturbations could be selected to also satisfy $k_z z_d>>1$ (although it would not be necessary be design). This might be equivalent to $k_z z>>1$,  in the case of a flattened logarithmic potential $\Phi_0\propto \ln(R^2+z^2/q^2)$, for example, or $k_z h>>1$ in the case of an exponential vertical distribution with scale length $h$.  }

In practice, finite (WKB) wave perturbations are described in what follows by introducing a truncation at some height $h_1$, above which the density becomes zero.  (Note that, beyond $h$, $\rho_1$ is still assumed to be considerably less than $\rho_0$.)  Such finite WKB perturbations are maximally flexible as they require only $k_z z_p>>1$ and not $k_z z_d>>1$ or $k_z h>>1$.  

Introducing a truncation in the perturbation at some height $h_1$ comes with one important additional requirement.  
To make the perturbation physical, it must  satisfy both Poisson's equation and Laplace's equation beyond $h_1$.  This introduces a strict boundary condition at the interface $\vert z\vert=h_1$, which places restrictions on the relationship between the vertical and radial perturbations.  This condition is determined below by matching 
the solution to Poisson's equation
with the solution to Laplace's equation at $z=h_1$ and requiring a similar matching of the gravitational force $\partial\Phi_1/\partial z$ at the interface (so that the gravitational force remains smooth).  Here $h_1$ is taken to be the disk scale height or greater.  

To satisfy Laplace's equation 

\begin{equation}
-k^2\Phi_1-\frac{m^2}{R^2}\Phi_1+\frac{\partial^2\Phi_1}{\partial z^2}=0 
\end{equation}
above and below the perturbation (at and beyond the vertical extent), the solution must have $k_z$=$i k_{\textrm{plane}}$.  Outside the perturbed part of the disk the potential thus becomes a decaying function $\Phi_1$$\propto e^{-\vert k_{\textrm{plane}} z\vert}$.
 
Meanwhile, over the extent of the perturbation, Poisson's equation 

\begin{equation}
-k_{\textrm{plane}}^2\Phi_1+\frac{\partial^2\Phi_1}{\partial z^2}=4\pi G\rho_1
\end{equation}
implies that
\begin{equation}
\Phi_1=\frac{-4\pi G\rho_1}{k_{\textrm{plane}}^2+k_z^2}\label{eq:pertpotential}
\end{equation}
in the WKB approximation, where again $k_{\textrm{plane}}^2=k^2+m^2/r^2$ and now $\partial^2\Phi_1/\partial z^2\approx -k_z^2\Phi_1$.  For Lin-Shu perturbations that are confined to an infinitssimally thin sheet $\Phi_1=-2\pi G\Sigma_1/\vert k_\textrm{plane}\vert$, in contrast \citep{toomre}.  

Below both even and odd density and potential wave perturbations are considered, although even perturbations are the focus of the remainder of the paper. (Odd wave perturbations can be shown to yield a consistent view of the main features of 3D disk instability.)  As discussed later in $\S$~\ref{sec:largeh1}, the 2D Lin-Shu dispersion relation and Toomre criterion are retrieved adopting an even wave perturbation, which can be envisioned as an over-density in the galaxy mid-plane.  This is the nominal perturbation for describing, i.e., the propagation of density waves in the disk.  The amplitudes of all perturbations considered in this work (whether even or odd) are assumed to be even functions of distance from the mid-plane.  Perturbations with amplitudes that are odd functions of $z$ have been shown to be stable by \citetalias{GLB}.  \\

\noindent\underline{Even WKB Perturbations}\\
\vspace*{-.1in}

In the case that the potential perturbation in the disk is an even function $\Phi_1\propto\cos{k_zz}$ and symmetric about the mid-plane with extent $h_1$, we obtain the following matching conditions

\begin{eqnarray}
A e^{-k_{\textrm{plane}}h_1}&=&B cos(k_z h_1)\nonumber\\
-k_{\textrm{plane}} A e^{-k_{\textrm{plane}}h_1}&=&-k_z B sin(k_z h_1)\nonumber
\end{eqnarray}
at the interface $h_1$, where $A$ is the amplitude of the potential perturbation beyond $\vert z\vert=h_1$ and $B$ is the amplitude within $h_1$.  %$\vert z\vert<h_1$. 
These conditions yields the relation \citep[see also][]{grivgedalin}:

\begin{equation}
\arctan{\frac{k_{\textrm{plane}}}{k_z}} = k_z h_1 \label{eq:klinkeven}
\end{equation}

In the standard long wavelength scenario with 
$k_{\textrm{plane}}<<k_z$, eq.(\ref{eq:klinkeven}) reduces to $k_{\textrm{plane}}/k_z\approx k_z h_1$.   Taking $h_1$=$h$, this yields the approximation 

\begin{equation}
\frac{\rho_0 k_{\textrm{plane}}^2}{k_{\textrm{plane}}^2+k_z^2}\approx\frac{\Sigma_0 k_{\textrm{plane}}}{2(1+k_{\textrm{plane}}h)}
\end{equation}
in terms of the unperturbed disk volume density $\rho_0$ and surface density $\Sigma_0\approx 2h\rho_0$.
This approximation is analogous to corrections incorporated into 2D dispersion relations (of single or multi-component disks) to account for weakened self-gravity when finite thickness is assumed \citep{toomre,vandervoort,jogsolomon,romeo92,ylin05,elm11}.  

This paper is also interested in the short wavelength regime where the above approximation is not valid. Here `short' is used in relation the vertical wavelength, not the disk scale height.  (Instability is indeed still restricted below the Jeans length.)  Most relevant in this scenario is the extent of the perturbation $h_1$.  In the limit $k_{\textrm{plane}}>>k_z$, the boundary condition in eq. (\ref{eq:klinkeven}) requires $k_z\approx(\pi/2) h_1^{-1}$ to lowest order, or that $\lambda_{R}$$<<$$\lambda_z\approx h_1$.  This scenario is consistent with $\lambda_z > h$ 
when we envision the perturbation's vertical edges at $h_1>h$, and it can thus be used to probe the instability regime in which $\lambda_R$ is brought down near the size of disk scale height. \\

\noindent\underline{Odd WKB Perturbations}\\
\vspace*{-.1in}

In the case that the potential perturbation in the disk is an odd function, $\Phi_1\propto\sin{k_zz}$, the matching conditions at $h_1$ above the plane become

\begin{eqnarray}
A e^{-k_{\textrm{plane}}h_1}&=&B sin(k_z h_1)\nonumber\\
-k_{\textrm{plane}} A e^{-k_{\textrm{plane}}h_1}&=&k_z B cos(k_z h_1)\nonumber
\end{eqnarray}
requiring that
\begin{equation}
\arctan{\frac{-k_z}{k_{\textrm{plane}}}} = k_z h_1. \label{eq:klinkodd1}
\end{equation}

Similarly, below the plane at -$h_1$, the boundary condition requires
\begin{equation}
\arctan{\frac{k_z}{k_{\textrm{plane}}}} = k_z h_1. \label{eq:klinkodd1}
\end{equation}

Thus in the long wavelength scenario $\vert k_{\textrm{plane}}\vert<<\vert k_z\vert$, allowed perturbations have a vertical wavenumber $k_z\sim(\pi/2)h_1^{-1}$ and radial wavenumbers are restricted to $\vert k_{\textrm{plane}}\vert$$<<$$1/h_1$ or $\vert k_{\textrm{plane}}\vert$$<$$1/h$ when $h_1>h$.  Perturbations in the short-wavelength limit $\vert k_{\textrm{plane}}\vert>>\vert k_z\vert$ have $\vert k_{\textrm{plane}}\vert\approx1/h_1$ and $k_z<<1/h_1$, which again can correspond to the case $\vert kh\vert\lesssim1$ and $\vert k_zh\vert<<1$ where $h_1>h$.  \\

\subsection{Obtaining the Conditions for Stability in 3D}\label{sec:equationsofmotion}
\subsubsection{Overview}
This section introduces the 3D perturbations from the previous section (corresponding to infinite non-periodic perturbations, infinite waves and finite WKB waves) into the linearized 3D equations of motion to solve for the perturbed motions in the radial, azimuthal and vertical directions.  The 3D dispersion relation is then obtained using the continuity equation, which couples these motions to the time evolution of the perturbed density.  
 
Using either the 3D version of the dispersion relation derived below ($\S$~\ref{sec:3ddispersionRelation}) or the 2D version ($\S$~\ref{sec:2dstability}), the conditions for stability can be easily determined
according to the expectation that 
a stable, non-growing mode (with Real $\omega$) must have $\omega^2$$>$0.  (Thus the line of stability is usually taken to be $\omega^2$=0; e.g. \citealt{BT})  In the interest of diagnosing the basic stability of disks to 3D perturbations, sections~\ref{sec:3ddispersionRelation} and on examine in detail the axisymmetric scenario with $m$=0, in which case $k_\textrm{plane}=k$. The calculations presented in what immediately follows, though, adopt an arbitrary $m$.  

\subsubsection{Motions in the Plane and in the Vertical Direction}\label{sec:motions}
To describe motions in our rotating gas disk, we adopt the Euler equations of motion in cylindrical coordinates, with $z$ oriented parallel to the axis of rotation.  We then introduce a small perturbation.  
Writing all quantities as the sum of perturbed and small unperturbed components (i.e. $\rho=\rho_{0}+\epsilon\rho_{1}$ and $v_R=v_{R,0}+\epsilon v_{R,1}$, etc., where $\epsilon$ is small) and keeping only terms to first order in small quantities, the linearized versions of the equations of motion are obtained \citep[see][]{BT}.  These are satisfied in this work by the perturbations introduced in $\S$~\ref{sec:perturbations} with the form $\Phi_1(R,\phi,z,t)=\Phi_a(R,z) e^{i(m\phi-\omega t)}$ where 
$\Phi_a(R,z)=\mathcal{F}(R,z) e^{ikR+ik_z z}$ and 
the radial gradient of $\mathcal{F}(R,z)$ is neglected in the WKB approximation.  Through Poisson's equation the density perturbation has a similar dependence, i.e. $\rho_{1}=\rho_{a}(R,z) e^{i(m\phi-\omega t)}$ where $\rho_{a}(R,z)=\mathcal{R}(R,z)e^{ikR+ik_z z}$.  Solutions to the linearized equations of motion (eq. (\ref{eq:EOM})) thus also have the form $v_{R,1}=v_{R,a}(R,z) e^{i(m\phi-\omega t)}$, $v_{\phi,1}=v_{\phi,a}(R,z) e^{i(m\phi-\omega t)}$ and $v_{z,1}=v_{z,a}(R,z) e^{i(m\phi-\omega t)}$ where $v_{R,a}(R,z)$, $v_{R,a}(R,z)$ and $v_{r,a}(R,z)$ are all $\propto e^{ikR+ik_z z}$.

Substituting the density and potential perturbations into the perturbed radial and azimuthal equations of motion, it can be shown \citep[adopting the convention in][]{BT} that
\begin{eqnarray}%\left[
v_{R,a}&=&-\frac{(\Phi_a+\sigma^2\frac{\rho_a}{\rho_0})}{\Delta}
\left(k(\omega-m\Omega)+i\frac{2m\Omega}{R}\right)\nonumber\\
&-&v_{z,a}\frac{2\Omega}{\Delta} \frac{dV_c}{dz}\label{eq:radvelocity}
\end{eqnarray}
\begin{eqnarray}
v_{\phi,a}&=&-\frac{(\Phi_a+\sigma^2\frac{\rho_a}{\rho_0})}{\Delta}\left(2Bik+\frac{m(\omega-m\Omega)}{R}\right)\nonumber\\
&+&iv_{z,a}\frac{(\omega-m\Omega)}{\Delta}\frac{dV_c}{dz}\label{eq:phivelocity}
\end{eqnarray}
where $V_c=\Omega R$ and 
\begin{eqnarray}
B&=&-\Omega-\frac{1}{2}R\frac{d\Omega}{dR}\\\nonumber
\Delta&=&\kappa^2-(m\Omega-\omega)^2\\\nonumber
\kappa&=&-4B\Omega.\nonumber
\end{eqnarray}

In the vertical direction, the linearized equation of motion 

\begin{equation}
i(-\omega+m\Omega)v_{z,1}=-\nabla\Phi_1-\sigma^2\left(\frac{\nabla\rho_1}{\rho_0}\right)
\end{equation}
implies
\begin{equation}
v_{z,a}=-\frac{k_z(\Phi_a+\sigma^2\frac{\rho_a}{\rho_0})}{(m\Omega-\omega)}+i\frac{(\nabla \mathcal{F}+\sigma^2\frac{\nabla \mathcal{R}}{\rho_0})}{(m\Omega-\omega)}e^{ikR+ik_z z}.
\label{eq:vertvelocity}
\end{equation}
where, at this stage, no assumption has been made about the relative sizes of the vertical perturbation's phase and amplitude variations.  Different choices for $k_z$ in relation to the perturbation amplitude and the unperturbed disk will be examined later in this work. 

These expressions for $v_{r,a}$, $v_{\phi,a}$ and $v_{z,a}$ are based on the assumption of equilibrium in the unperturbed disk, such that $v_{r,0}$=0, $v_{z,0}$=0 and $v_{\phi,0}$$\approx(R d\Phi_0/dR)^{1/2}=\Omega R$, neglecting the pressure term $(\partial p_{0,\phi}/\partial R)/\rho_0$ since $\sigma_{\phi,0}$$<<$$\Omega R$.  

Adopting the WKB approximation in the radial direction leads to further simplification.  This work focuses on scenarios in which the radial variation in the amplitude of the potential perturbation is comparable to (and no less than) the radial gradient in the unperturbed disk (as discussed in \ref{sec:perturbations}).  As is typical, then, the factors proportional to $1/R$ in eqs. (\ref{eq:radvelocity}) and (\ref{eq:phivelocity}) are neglected relative to those that are proportional to $k$ \citep[e.g.][]{BT}. The WKB condition $k R_p>>1$ ($\S$~\ref{sec:perturbations}) is satisfied by $k R>>$1 assuming $k$ increases towards small $R$. 

The terms proportional to $v_{z,1}$ in the expressions for $v_{r,1}$ and $v_{\phi,1}$ are similarly neglected in the set-up of interest here, since the rotational lag

\begin{equation}
\frac{dV_c}{dz}\approx\frac{1}{2\Omega}\frac{d}{dz}\frac{d}{dR}\Phi_{0}
\end{equation}
(again assuming that the radial pressure gradient is negligible) 
contains a factor considerably smaller than $k\Phi_a$.\footnote{Appendix \ref{sec:rotationallagappendix} identifies the precise set of perturbations for which the lag term is negligible. }. 

With an identical vertical perturbation specifically in the WKB approximation, \cite{grivgedalin} arrive at a different expression for the perturbed vertical velocity $v_{z,1}$, as the disk in their scenario of interest is out of hydrostatic equilibrium.  This introduces factors proportional to $v_{z,0}$, such that the numerator in eq. (\ref{eq:vertvelocity}) includes include a term proportional to $\nu$, the vertical epicyclic frequency.   

\subsection{The 3D Dispersion Relation}\label{sec:3ddispersionRelation}
Next we consider the perturbed continuity equation in cylindrical coordinates

\begin{equation}
i(m\Omega-\omega)\rho_1+\frac{1}{R}\frac{d}{d R}(R\rho_0v_{R,1})+\frac{im\rho_0}{R}v_{\theta,1}+\frac{d}{dz} (\rho_0 v_{z,1})=0,\label{eq:fullcontinuity}
\end{equation}
including the vertical term, using the fact that $v_{z,0}$=0 for the continuity-obeying equilibrium unperturbed disk and keeping only terms lowest order in the perturbation.  

Adopting the WKB approximation with the assumption that $kR>>1$ \footnote{This assumption is weakened in Appendix \ref{sec:nonaxisym} to examine the conditions for stability in the presence of perturbations that are non-axisymmetric in the plane. } 
leads to the simplification %.  

\begin{equation}
i(m\Omega-\omega)\rho_1+\rho_0\frac{\partial v_{R,1}}{\partial R}+\rho_0\frac{\partial v_{z,1}}{\partial z}+v_{z,1}\frac{\partial \rho_0}{\partial z}=0.\label{eq:3ddispersion}%+k_z\rho_1v_{z,0}
%\frac{\partial \rho_1}{\partial t}+k\rho_0v_{r,a}+k_z\rho_0v_{z,a}=0%+k_z\rho_1v_{z,0}
\end{equation}
(The $v_{\phi,1}$ term is small compared to the other two velocity terms in the WKB approximation and is dropped.) 

Before considering the generic case in which $k_z\neq0$ and $k\neq0$ (in section \ref{sec:3dmidplane}), below we will considered radial and vertical perturbations separately.  \\

\subsubsection{Vertical-only Perturbations ($k_z\neq$0, $k$=0)}\label{sec:verticalonly}
Now taking $z_d=(\partial \ln\rho_0/\partial z)^{-1}$, 
substituting in the expression for $v_{z,1}$ and setting $k$=0, eq. (\ref{eq:3ddispersion}) becomes

\begin{eqnarray}
0&=&(m\Omega-\omega)^2\rho_1\nonumber\\
&-&k_z^2\left(\Phi_1\rho_0+\sigma^2\rho_1\right)%\nonumber\\%+k_z\rho_1v_{z,0}
+i k_z\Phi_1\rho_0\left(\frac{1}{z_d}\right)\nonumber\\
&+&\mathcal{A}e^{i(m\Omega-\omega)t}\label{eq:vertonly2}
%\frac{\partial \rho_1}{\partial t}+k\rho_0v_{r,a}+k_z\rho_0v_{z,a}=0%+k_z\rho_1v_{z,0}
\end{eqnarray}
where 
\begin{eqnarray}
\mathcal{A}&=&e^{ikr+ik_zz}\Big[\rho_0\nabla^2\mathcal{F}+\sigma^2\nabla^2\mathcal{R}\nonumber\\
&+&\left(\frac{1}{z_d}\right)\rho_0\nabla \mathcal{F}\nonumber\\
%&+&\left(\frac{1}{z_d}\right)\left(\nabla F+\sigma^2\frac{\nabla p}{\rho_0}\right)\nonumber\\
&+&2ik_z\left(\rho_0\nabla \mathcal{F}+\sigma^2\nabla \mathcal{R}\right)\Big]
\end{eqnarray}

In the non-wave scenario ($k_z=0$) the dispersion relation reads
\begin{eqnarray}
0&=&(m\Omega-\omega)^2\rho_1+\Big[\rho_0\nabla^2\mathcal{F}+\sigma^2\nabla^2\mathcal{R}\nonumber\\
&+&\left(\frac{1}{z_d}\right)\rho_0\nabla \mathcal{F}\Big]e^{ikr+i(m\Omega-\omega)t}
%&+&\left(\frac{1}{z_d}\right)\left(\nabla F+\sigma^2\frac{\nabla p}{\rho_0}\right)\Big]e^{ikr+i(m\Omega-\omega)t}
\label{eq:vertGLB}
\end{eqnarray}
whereas in a WKB scenario, 
\begin{eqnarray}
0&=&(m\Omega-\omega)^2\rho_1-k_z^2\left(\Phi_1\rho_0+\sigma^2\rho_1\right)\nonumber\\
&+&i k_z\Phi_1\rho_0\left(\frac{1}{z_d}\right)\label{eq:vertWKB}
\end{eqnarray}

The imaginary, out-of-phase term in eq. (\ref{eq:vertWKB}) is notably negligible when $k_z z_d>>1$.   
This would be equivalent to the condition required to keep an infinite perturbation consistent with WKB approximation (using the requirement $z_p$=$z_d$ to keep the perturbation small with respect to the unperturbed disk as $\vert z\vert~\rightarrow~\infty$).  Likewise, the second factor in the term in square brackets in eq. (\ref{eq:vertGLB}) drops when $k_z z_d>>1$, considerably simplifying the expression.   %, which is not invoked here.  
However, in the case of the unperturbed Gaussian vertical profile (for which $1/z_d=z/h^2$), $k_z z_d>>1$ applies only very near the galactic mid-plane, i.e. where $z<<(k_z h)h$.  Thus, both the in-phase and out-of-phase terms are relevant for the overall evolution of extended perturbations.     

Indeed, in the special case highlighted in the next section in which the perturbation extends to $\pm$ infinity and tracks the decrease in $\rho_0$ with increasing $z$ (as in our equilibrium disks) then integration over the vertical direction from $-\infty$ to $\infty$ yields zero when all terms in eqs. (\ref{eq:vertonly2}), (\ref{eq:vertGLB}) and (\ref{eq:vertWKB}) are included.   In other words, $\rho_0v_{z,1}\vert_{-\infty}^{\infty}=0$. This is the `no mass flux at infinity' requirement invoked by GLB.  As a result, the vertical-only 2D dispersion relation in this 'no mass flux' scenario reads $(m\Omega-\omega)^2=0$ signifying that the vertical direction is neutrally stable to infinite axisymmetric perturbations and stable to all non-axisymmetric perturbations (since $\omega^2=m^2\Omega_p^2$ when $k$=0).  

This neutral stability characteristic of the vertical direction is leveraged when calculating the 2D dispersion relation later in $\S$~\ref{sec:2dstability}, following GLB.  Below, it will first be useful to examine how the terms in eqs. (\ref{eq:vertGLB}) and (\ref{eq:vertWKB}) proportional to $1/z_d$ contribute to this neutral vertical stability.  \\

\noindent\underline{Stability Away from the Mid-plane}\\
\vspace*{-.1in}

In the case of the non-wave perturbation, the third term in eq. (\ref{eq:vertGLB}) dominates away from the mid-plane and

\begin{equation}
(-\omega+m\Omega)^2\rho_1\approx-\frac{\rho_0}{z_d}\nabla \Phi_1
\end{equation}
where $\Phi_1=\mathcal{F}$.  This corresponds to stability ($\omega^2>0$) everywhere since the right-hand side is always positive when $\nabla \Phi_1>0$ and $z_d<0$, as it is in the equilibrium disks under consideration. 

Stablility far above the mid-plane is also a feature of periodic wave perturbations. Consider eq. (\ref{eq:vertWKB}) in a scenario in which the perturbation is extended but finite, for example.\footnote{The perturbation must be finite or it will not satisfy the WKB approximation assumed for the present exercise. This requirement is not invoked in other sections unless noted. } 
In the regime $k_z z_d<<1$, or at heights much larger than $h$ (far away from the mid-plane),  the vertical-only continuity equation reads

\begin{equation}
-\frac{\omega^2 (k^2+k_z^2-T^2)h^2}{f_g\nu^2k_z z}=\textrm{cot}(k_z z+kr+(m\Omega-\omega)t)
\end{equation}
adopting our Gaussian vertical density profile and letting $\nu^2=4\pi G\rho_0 f_g$.  
Since the arccotangent of the left hand side is $\pm\pi/2$ for all $z>>h$, $\omega$ is always real.  It remains real as $z$ approaches nearer to $h$, where the arccotangent of the left hand side is a small positive or negative quantity.  \\

\noindent\underline{(In)stability Near the Mid-plane}\\
\vspace*{-.1in}

The stability away from the mid-plane is in contrast to the situation very near the mid-plane.  In the case of wave perturbations, the dispersion relation where $k_z z_d>>1$ (and adopting $m$=0 for simplicity) becomes:  

\begin{equation}
\omega^2=-4\pi G\rho_0\frac{k_z^2}{k_z^2-T^2}+k_z^2\sigma^2\label{eq:3dmid}
\end{equation}
or

\begin{equation}
\omega^2=-4\pi G\rho_0+k_z^2\sigma^2\label{eq:3dmid}
\end{equation}
in the limit $k_z>>T$.  (The stability of non-wave perturbations near the mid-plane is examined in $\S$~\ref{sec:3dmidplane}.)
In this situation, perturbations have the opportunity for growth as long as $k_z <(4\pi G\rho_0)/\sigma_z^2=k_J$.  In other words, very near to the roughly constant-density mid-plane, 
instability in the vertical direction proceeds in a Jeans-like manner, unaffected by rotation \citep{chandrasekhar54} and restricted to similar scales.  \\

\noindent\underline{Total Neutral Stability}\\
\vspace*{-.1in}

As exemplified by the `no-mass-flux at infinity' case described above and considered in detail in $\S\S$ \ref{sec:infGLB} and \ref{sec:infwave}, the combination of  instability near the plane with stability away from the plane results in a neutrally-stable disk.\footnote{It is worth noting that the vertical variation in $\omega$ discussed here has not been made explicit in writing eqs. (\ref{eq:vertonly2}), (\ref{eq:vertGLB}) and (\ref{eq:vertWKB}).  This corresponds to the assumption that the perturbation's amplitude and/or phase variations are faster, i.e. $\vert T\vert >>(d\omega/dz) t$ or $\vert k_z\vert >>(d\omega/dz) t$ everwhere.  } The disk is thus neither as unstable as predicted where $k_z z_d>>1$ or as stable as predicted where $k_z z_d<<1$. 

Still, eq. (\ref{eq:3dmid}) does suggest that an avenue to avoid stability would be to perturb the disk over a limited extent, very near the mid-plane.  
As demonstrated later in $\S$ \ref{sec:finWKB}, perturbations extending a finite distance above and below $z=0$ are defined by the non-zero mass flux they entail over their extent, with the consequence that the 2D dispersion relation retains terms associated with the vertical direction.  As the perturbation's height decreases, stability approaches the behavior predicted in the limit $k_z z_d>>1$ near the mid-plane. 

\subsubsection{Radial-only Perturbations ($k\neq$0, $k_z$=0)}\label{sec:radialonly}
\indent Setting $m$=0 and $k_z$=0 in eq. (\ref{eq:3ddispersion}) and substituting in the expression for $v_{r,1}$, the axisymmetric radial-only dispersion relation reads 

\begin{equation}
\omega^2=\kappa^2-4\pi G\rho_0\frac{k^2}{k^2-T^2} +\sigma_r^2k^2\label{eq:raddisp}\\
\end{equation}
or 
\begin{equation}
\omega^2=\kappa^2-4\pi G\rho_0 +\sigma_r^2k^2\label{eq:raddisp}\\
\end{equation}
in the limit that $k>>T$. 
This is a restatement of the finding that wave perturbations perpendicular to the axis of rotation (in this case, the radial direction) can be stabilized by rotation, since the scales of instability are pushed over the Jeans length.  This was first found by \cite{chandrasekhar54} in the case of uniform rotation, then generalized by \citet{BelSchatzman} for non-uniform rotation (as considered here) and then confirmed to apply in the presence of vertical flattening \citep{safronov}.   

This scenario resembles the case of `no vertical mass flux at infinity' perturbations and 3D perturbations near the mid-plane and so a discussion of the instability scale is postponed until $\S\S$~\ref{sec:3dmidplane} and~\ref{sec:infGLB} .  For now it should be noted that simply omitting a vertical perturbation very clearly does not retrieve the 2D Lin-Shu dispersion relation.\footnote{As discussed later in $\S$~\ref{sec:largeh1}, to retrieve the Lin-Shu dispersion relation in the long-wavelength limit starting from the 3D dispersion relation requires taking the limit in which the disk is an infinitesimally thin sheet with $\sigma_z\rightarrow 0$).  
}  Instabilities instead have a Jeans-like quality 
even in the presence of rotation.  (Eq. [\ref{eq:raddisp}] indeed approaches the condition for Jeans instability in the limit $\kappa\rightarrow0$.)

\cite{jogsolomon} pointed out this resemblance to Jeans instability by taking the 2D dispersion relation (see eq. (\ref{eq:linshu}), $\S$~\ref{sec:largeh1}) in the small wavelength (large $k$) limit, opposite to the standard long wavelength regime.  As examined in $\S$~\ref{sec:3dmidplane} (and later in $\S$~\ref{sec:smallh1}), this 3D quality signifies a change in the disk stability threshold compared to the value required for stability to Lin-Shu density-wave perturbations.  

\subsubsection{An Assessment of 3D Stability Near the Mid-plane}\label{sec:3dmidplane}
This section describes stability and fragmentation from a fully 3D perspective embedded within the gas disk, near to the galactic mid-plane.  Later this view is traded for a 2D perspective that can be used to assess the overall stability of the disk (including all material out to $z=\pm\infty$).  

Including both radial and vertical perturbations (with $k_z\neq0$ and $k\neq0$) and substituting in the expression for $v_{r,1}$, equation (\ref{eq:3ddispersion}) can be rewritten as 
\begin{eqnarray}
&\rho_1&\bigg[(m\Omega-\omega)+\left(-\frac{4\pi G\rho_0}{k_\textrm{plane}^2+k_z^2-T^2}+\sigma_r^2\right)\frac{k^2(m\Omega-\omega)}{\Delta}\nonumber\\
&+&\frac{C_z}{(m\Omega-\omega)}\bigg]=0, \label{eq:3dstep1}
\end{eqnarray}
where $C_z$ represents vertical stability and is equated with either the second term on the right hand side of eq. (\ref{eq:vertGLB}) or the sum of last two terms on the right side of eq. (\ref{eq:vertWKB}), specifically including both in-phase and out-of-phase terms.  Notice that when $C_z$ is positive (negative) in eqs. (\ref{eq:vertGLB}) or (\ref{eq:vertWKB}) the vertical direction is unstable (stable).  

The 3D dispersion relation in eq.(\ref{eq:3dstep1}) is quadratic in $\omega^2$, with solutions

\begin{equation}
\omega^2=\frac{\omega_{min}^2}{2}\left(1\pm\sqrt{1+\frac{4C_z\kappa^2}{\omega_{min}^4}}\right)
\end{equation}
where

\begin{equation}
\omega_{min}^2=\kappa^2+\left(\frac{-4\pi G \rho}{k^2+k_z^2-T^2}+\sigma_r^2\right)k^2
-C_z  
\end{equation}
in the case that $m$=0.
It is straightforward to show that the condition $\omega^2$$<$0 can be met when both

\begin{equation}
\omega_{min}^2<0  \label{eq:3dstability}
\end{equation}
and $C_z>0$, corresponding to vertical instability.  
(When $\omega_{min}^2$ is positive, there is a limited range of conditions under which one of two branches of $\omega^2$ can still become negative.  But we neglect such a scenario here, considering that the criterion in eq. (\ref{eq:3dstability}) is readily met.)

Notice that when $\omega_{min}^2<0$ and $C_z>0$, then $\omega_{min}^2$ is the minimum that $\omega^2$ can reach.  
In what follows, eq. (\ref{eq:3dstability}) is used as the condition for instability, with the understanding that growth may happen faster than indicated by $\omega_{min}$.  Below, conditions on $k$ (and/or $k_z$) for instability specifically near the mid-plane are obtained from eq. (\ref{eq:3dstability}) in the case of both wave and non-wave 3D perturbations. \\

\noindent\underline{Wave Perturbations}\\
\vspace*{-.1in}

For wave perturbations near the mid-plane (i.e. in the limit $k_z z_d>>1$) that are also assumed to locally satisfy the WKB approximation ($k_z>>T$) for illustration purposes, the 3D dispersion relation is written

\begin{equation}
\kappa^2+\left(\frac{-4\pi G \rho_c}{k^2+k_z^2}+\sigma_r^2\right)k^2
+\left(\frac{-4\pi G \rho_c}{k^2+k_z^2}
+\sigma_z^2\right)k_z^2<0.  \label{eq:3dstabilitywave}
\end{equation}
 using that 
\begin{equation}
C_z=
-k_z^2\left(-\frac{4\pi G\rho_c}{k_\textrm{plane}^2+k_z^2}+\sigma_z^2\right)\label{eq:czmidplane}
\end{equation}
in this limit (see previous section) with $\rho_c=\rho(z\rightarrow 0)$.

Now, setting $k_S^2=k^2+k_z^2$ (with $S$ denoting 'shell'), instability is found to require 

\begin{equation}
\kappa^2-4\pi G \rho+\sigma^2k_S^2+k_z^2(\sigma_z^2-\sigma_r^2)<0 \label{eq:nearshellmidplane}
\end{equation}
or 
\begin{equation}
k_S^2<k_J^2\left(1-\frac{\kappa^2}{4\pi G\rho_c}\right)
\end{equation}
assuming that the velocity dispersion is isotropic ($\sigma_z$=$\sigma_r$=$\sigma$). 
Stability within the roughly constant density region staddling the mid-plane 
thus takes on a Jeans-like quality, though rotation succeeds in increasing the size of stable fragments.  

Rotation can also eventually suppress instabilities above a threshold 

\begin{equation}
Q_M\equiv\kappa^2/(4\pi G\rho_c)>1.
\end{equation}  
It is notable that the form of this threshold resembles the 3D threshold $\kappa^2/(\pi G\rho_c)\approx 0.3$ determined for the overall disk by GLB better than it matches $Q_T$.  As discussed in detail later in $\S$~\ref{sec:infGLB}, the difference in the numerical value of the threshold is a consequence of the vertical extent of the perturbed region.  

The threshold $Q_M=1$ also corresponds to higher stability threshold than $Q_T=1$.  In the case of weakly self-gravitating disks (with $\Sigma=\rho_c\sqrt{2\pi}h$), 

\begin{equation}
Q_M=\frac{\pi\alpha^2f_gQ_T^2}{8}, \label{eq:QMQT}
\end{equation}
while in the case of fully self-gravitating disks (with $\Sigma=\rho_c2h$)

\begin{equation}
Q_M=\frac{\alpha^2f_gQ_T^2}{4}.   
\end{equation}
Thus, $Q_M=1$ is equivalent to $Q_T\approx 2$, signifying that disks are more susceptible to partial 3D instability (endemic to the mid-plane) than to total destablization described by the 2D Toomre criterion, as discussed more in $\S$~\ref{sec:growthrates}.

It is also noteworthy that, as a criterion specifically on the radial $k$ wavenumber, eq. (\ref{eq:3dstabilitywave}) implies 

\begin{equation}
k^2<k_J^2\left(1-Q_M-k_z^2h^2\right)\label{eq:midplane3Djeans}
\end{equation}
with a stability threshold $Q_M=1-k_z^2h^2$.  Here the radial Jeans length $k_J=4\pi G\rho/\sigma_r^2$.  Since $h$ is roughly equivalent to the effective Jeans length (applicable in the presence of thermal and non-thermal motion), it is only when the disk is perturbed on scales larger than the Jeans length that radial fragmentation is seeded.  That is, the largest perturbations,  with $k_zh<<1$, correspond to the highest threshold and thus most easily seed radial fragmentation.  \\ 

From a more qualitative perspective, the onset of this `mid-plane' Jeans-like instability can be described as follows:
At the mid-plane where the density is approximately constant, gas pressure applies a negligible force and only the perturbed pressure force is left to compete with self-gravity.  The vertical component of this force is negligible when the wavelength of the perturbation is large, i.e. $k_zh<1$ or when the disk is perturbed above the vertical (effective) Jeans length.  As a result, the primary competition against self-gravity comes from the pressure force in the plane.  Since the pressure force is scale-dependent while self-gravity at the mid-plane is not, the result is that the disk is able to destabilize, but only on scales larger than the radial Jeans length (lengthened by rotation).  

It is worth noting that, although this instability is described as occurring `at the mid-plane', it is limited by pressure to scales larger than the vertical Jeans length.  Thus the scale height sets the minimum vertical extent of the region that becomes unstable.  Indeed, in either the radial or vertical directions, the disk is stabilized by pressure below the Jeans length.  According to eq. (\ref{eq:midplane3Djeans}), rotation also contributes to stability on the largest scales. \\

\noindent\underline{Non-wave Perturbations}\\
\vspace*{-.1in}

Non-wave ($k_z<<T$) vertical perturbations that are infinite (and satisfy $1/z_p=1/z_d$) exhibit almost identical behavior near the mid-plane.  For these, 

\begin{equation}
C_z=\rho_0\nabla_z^2\Phi_1+\sigma^2\nabla_z^2\rho_1+\left(\frac{1}{z_d}\right)\left(\nabla_z \mathcal{F}+\sigma^2\frac{\nabla_z \mathcal{R}}{\rho_0}\right)\label{eq:czmidnonwave}
\end{equation}
(see previous section).  In the limit, $z<<h$ where the unperturbed and perturbed densities are roughly constant and $1/z_d=z/h^2<<1/h$ (for our adopted Gaussian equilibrium vertical profile), the second and third (pressure) terms drop.
Thus, substituting eq. (\ref{eq:czmidnonwave}) into eq. (\ref{eq:3dstep1}), the condition for instability becomes

\begin{equation}
\kappa^2+\frac{\rho_0}{\rho_1}k^2\Phi_1+\sigma_r^2k^2
-\frac{\rho_0}{\rho_1}\nabla_z^2\Phi_1<0
 \label{eq:3dstabilityGLBmidv0}
 \end{equation}
 or
\begin{equation}
\kappa^2-4\pi G \rho_0+\sigma_r^2k^2<0
 \label{eq:3dstabilityGLBmid}
 \end{equation}
using Poisson's equation for the substitution $\nabla_z^2\Phi_1=4\pi G\rho_1+k^2\Phi_1$.  
Specifically near the mid-plane, instability in the presence of an arbitrary infinite perturbation is possible as long as 

\begin{equation}
k^2<k_J^2(1-Q_M)
  \end{equation}
  with stability once again setting in above the threshold $Q_M=1$.  The minimum instability scale is thus the radial Jeans length in the radial direction and effectively the scale height in the vertical direction (or, more precisely, the extent of the region where the disk density is approximately constant).

As discussed in $\S$~\ref{sec:verticalonly}, the factors proportional to $1/z_d$ that were neglected here become important away from the mid-plane and serve to lower the threshold for the overall stability of the disk to infinitely extended perturbations.  This was previously determined by GLB, who calculated a total (non-wave) perturbation-weighted threshold $\bar{Q}_M<Q_M=1$ from the 2D dispersion relation derived by integrating over the vertical direction from $-\infty$ to $+\infty$. The next section examines this further, expanding the calculation to include the infinite and finite wave perturbations considered in this work.  

\section{2D Stability criteria}\label{sec:2dstability}
\subsection{Overview}
The 3D dispersion relation encodes the evolution of the perturbed density in the presence of the radial and vertical motions that develop in response to gravity, rotation and gas pressure.  
In the previous section, this evolution was shown to correspond to stability or growth in a manner that is sensitive to distance from the mid-plane ($\S$ \ref{sec:verticalonly}); 
near $z=0$, perturbations can be unstable above the radial and vertical Jeans lengths, while beyond $\vert z\vert \approx h$, the disk is characterized by stability.  This has two important implications.  First, the entire disk is neither as unstable (or stable) as predicted near (or far) from the mid-plane, and we can expect the overall stability threshold (determined from the 2D dispersion relation, after integration over the vertical direction) to be lower than $Q_M$=1 predicted near $z=0$.  Second, perturbations representing density enhancements with varying extents around on the mid-plane will have different stability thresholds, with the most confined perturbations best able to avoid the stability at locations far beyond $h$.

To examine these implications further, the following sections derive the 2D dispersion relation in a number of scenarios.  The first and second of these focus on the case of infinite non-wave and wave perturbations that satisfy the no-mass flux at infinity requirement.  Like the unperturbed disk density, these perturbations fall slowly to zero with increasing $z$ by setting $1/z_p\gtrsim 1/z_d$ where $z_d$ captures the gradient in the equilibrium density. % towards $\vert z\vert\rightarrow\infty$. 
(This also keeps them small with respect to $\rho_0$ everywhere.)  The infinite case is then compared with a scenario in which the perturbation is wave-like and allowed to have some finite extend $h_1$ above and below the mid-plane.    
As discussed earlier, these wave perturbations can be studied using the WKB approximation (assuming some arbitrary amplitude variation), since their truncation prevents them from violating the required $\rho_1/\rho_0<<1$ as $\rho_0\rightarrow 0$.  By examining these finite perturbations in two main regimes $h_1/h<<1$ and $h_1/h>>1$, bounds are placed on the possible range of stability thresholds that apply to 3D disks. 

For illustration purposes, in what follows the Gaussian vertical distribution is specifically adopted, although vertical integration in the case of the $sech^2$ profile is also discussed.  In addition, only even vertical perturbations are considered.  
As a diagnostic of stability in general, the case of axisymmetry in the plane ($m$=0) is specifically highlighted, although non-axisymmetry is considered in Appendix \ref{sec:nonaxisym}.  
%}

\subsection{Zero Vertical Mass Flux Infinite Non-wave (GLB) Perturbations}\label{sec:infGLB}
To serve as a reference for stability thresholds calculated in this work, this section presents a derivation of the threshold implied by the 2D dispersion relation in the scenario examined by GLB.  This involves a radial WKB wave perturbation and a generic infinite non-wave (non-periodic) vertical perturbation that satisfies the `no vertical mass flux at infinity' condition introduced by those authors.  

For the perturbations under consideration, Poisson's equation reads
\begin{equation}
(-k^2+T^2)\Phi_1=4\pi G\rho_1
\end{equation}
where $T$ measures the amplitude variation, defined in the previous section.  

These perturbations entail no mass flux at $z=\pm\infty$ when their amplitudes are even functions of $z$ and fall to zero as $\vert z\vert\rightarrow\infty$.  In practice this amplitude variation has to be faster than the vertical variation of the density in the unperturbed (equilibrium) disk, in order that it remains small at all locations ($\rho_1/\rho_0<<1$) .  In this case, the integral of the third (vertical) term in the continuity equation

\begin{equation}
\int_{-\infty}^{\infty}\frac{d(\rho_0v_{z,1})}{dz}dz=\rho_0v_{z,1}\vert_{-\infty}^{\infty}=0.
\end{equation}
For this scenario, the 2D dispersion relation obtained by vertical integration of the continuity equation becomes 
\begin{eqnarray}
&0&=\int_{-\infty}^{\infty}\Delta\rho_1dz-\int_{-\infty}^{\infty}\rho_0 \frac{4\pi G\rho_1}{k^2-T^2}k^2dz\nonumber\\&+&\int_{-\infty}^{\infty}\sigma_r^2\rho_1k^2dz+\int_{-\infty}^{\infty}\frac{\Delta}{(-\omega+m\Omega)}\frac{d (\rho_0 v_z)}{dz}dz
\end{eqnarray}
which can be written as 

\begin{equation}
\bar{\omega}^2=\kappa^2-\gamma_T 4\pi G\rho_c +\sigma_r^2k^2\label{eq:GLB2d}
\end{equation}
where the perturbation-weighted 

\begin{equation}
\bar{\omega}^2=\frac{\int_{-\infty}^{\infty}\omega^2\rho_1dz}{\int_{-\infty}^{\infty}\rho_1dz}
\end{equation}
and the factor

\begin{equation}
\gamma_T=\frac{\int_{-\infty}^{\infty}\frac{(4\pi G\rho_0)k^2}{k^2-T^2}\rho_1dz}{\int_{-\infty}^{\infty}\rho_1dz}.\label{eq:gammaInt}
\end{equation}
(Note that vertical variation in $\kappa^2$ is neglected here but is considered in Appendix \ref{sec:rotationallagappendix}.)

For the overall disk to become unstable, $\bar{\omega}^2$ must be less than zero. This translates into the instability condition 

\begin{equation}
k^2<\frac{4\pi G\bar{\rho}}{\sigma_r^2}(\sqrt{2}\gamma_T-\bar{Q}_M),
\end{equation}
which is associated with the stability threshold

\begin{equation}
\bar{Q}_M=\sqrt{2}\gamma_T
\end{equation}
in terms of the mean density 

\begin{equation}
\bar{\rho}=\frac{\int\rho_0^2 dz}{\int\rho_0 dz}=\frac{\rho_c}{\sqrt{2}}
\end{equation}
for the Gaussian vertical profile\footnote{  
(For the self-gravitating disk, $\bar{\rho}=2/3\rho_c$.)} and where $\bar{Q}_M=\kappa^2/(4\pi G\bar{\rho})$.  

\begin{figure}[t]
%\begin{flushleft}
%\vspace*{-.15in}
\begin{center}
\begin{tabular}{c}
\includegraphics[width=0.915\linewidth]{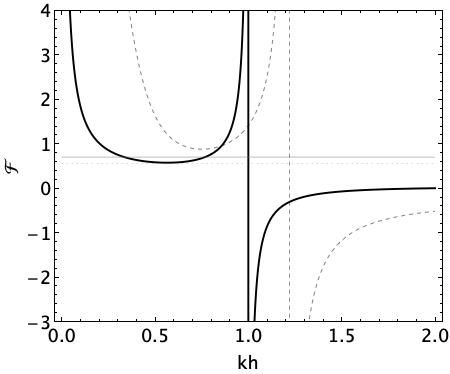}%&\includegraphics[width=0.495\linewidth]{ratesfg2.pdf}
\end{tabular}
\end{center}
%\vspace*{-.15in}
\caption{The behavior of $\mathscr{F}$ (or $1/(4\sqrt{2}\gamma_T)$ in terms of $\gamma_T$ introduced in the text) that sets the threshold on $\pi G\bar{\rho}/\kappa^2$ in the formalism of GLB and thus defines the onset of stability in 3D rotating flattenend disks.  The threshold calculated by GLB is shown in black. The approximation described in the text is shown as a black dashed line.  Two gray horizontal lines depict the minimum threshold value calculated numerically in this work ($\mathscr{F}=0.56$; dotted) and estimated by GLB ($\mathscr{F}=0.73$; solid).  Instability is possible whenever $\pi G\bar{\rho}/\kappa^2>\mathscr{F}$ or whenever $\bar{Q}_M<\sqrt{2}\gamma_T=1/(4\mathscr{F})$.  }
\label{fig:gamma}
%\end{flushleft}
\end{figure}

The quantity $1/(4\sqrt{2}\gamma_T)$ is equivalent to the function $\mathscr{F}$ evaluated analytically (with great effort) by GLB in the case of the fully self-gravitating disk.  (In their formalism, $\mathscr{F}$ sets the threshold on the quantity $\pi G\bar{\rho}/\kappa^2$.)  A few simplifying assumptions make it possible to perform the integral with greater transparency while still obtaining the main features of $\mathscr{F}$.  In the estimate below, the Gaussian profile (which applies to the idealized weakly self-gravitating case) is adopted (as opposed to assuming that the gas is self-gravitating) and the quantity $\nabla^2\Phi_1/\Phi_1=T^2$ is approximated as $2/z^2$ in the present case that $1/z_p\approx 1/z_d$.\footnote{From the perturbed vertical equation of motion, \begin{equation}
\nabla\Phi_1=-\frac{\sigma^2}{\rho_0}\nabla\rho_1+f(z)
\end{equation}
where $f(z)=-iv_{z,1}(-\omega+m\Omega)$, it can be shown that when $z_p=z_d$, 
\begin{equation}
T^2=\frac{\nabla^2\Phi_1}{\Phi_1}=\frac{2}{z^2}\frac{(1+\nabla f(z))}{\left(1+\frac{(2/z^2)\int f(z)dz}{4\pi G\rho_{1,c}}\right)}.
\end{equation} 
Below this is approximated as $2/z^2$, but the full expression for $T^2$ is handled in the derivation by GLB.
}
With these assumptions, 

\begin{equation}
\gamma_T\approx 1/\sqrt{2} - i \frac{e^{-2/(k_zh)^2} \sqrt{\pi}}{k_zh} - 
 \frac{2 \textrm{Dawson}\left(\frac{\sqrt{2}}{k_zh}\right)}{k_zh}.
\end{equation}
in terms of the Dawson integral

\begin{equation}
\textrm{Dawson}\left(y\right)=e^{-y^2}\int_0^y e^{t^2}dt.
\end{equation}
Although rough, this approximation brings us close to the result of GLB (see Figure \ref{fig:gamma}), mainly by capturing three main features: at small $k_z z$, the integrand in eq. (\ref{eq:gammaInt}) is proportional to $(k_z z)^2/2$ and negative, there is a singularity at $z=T=2/k_z$, and at large $k_z z$ the integrand is independent of $k_z$ and positive.  The similarity between $\mathscr{F}$ and $1/\gamma_T$ is also helped by the similarity between Gaussian and $\mathrm{sech^2}$ profiles generally, and especially near $z\approx 0$ where $T^2/k^2=2/(k z)^2$ is large.  

Following GLB, from $\mathscr{F}$ (or $1/(4\sqrt{2}\gamma_T)$) we can identify the following characteristics in the stability behavior of disks overall (out to $\pm\infty$): there are two critical regimes, $kh\rightarrow0$ and $kh\rightarrow1$, and a critical most-unstable wavenumber $kh\sim 0.5-0.6$ where the minimum stability threshold of $\mathscr{F}\approx 0.6$ is reached, corresponding to $\bar{Q}_M=\sqrt{2}Q_M\approx0.45$. \footnote{The estimates for stability above $\bar{Q}_M=0.45$ or below $\pi G\bar{\rho}/\kappa^2=0.56$ due to rotation were determined by GLB in the case of a fully self-gravitating isothermal disk.  (This is a recalculation of the threshold 0.73 determined by GLB, located by finding the minimum in their function $\mathscr{F}(m=kT)$ for the self-gravitating isothermal disk.) Note that, as determined by GLB, in the case of the steeper equation of state $P\propto\rho^2$, the threshold lowers to $\bar{Q}_M=0.27$ ($\pi G\bar{\rho}/\kappa^2=1.1$) and reduces still further to $\bar{Q}_M=0.14$ ($\pi G\bar{\rho}/\kappa^2=1.75$) for an incompressible disk. }  The sign change above $kh>1$ indicates that the disk is always stable in this regime. 

Since $\mathscr{F}$ (or $1/\gamma_T$) is relatively flat across the range $0\lesssim kh\lesssim 1$, in practice the condition for instability can be well approximated by

\begin{equation}
k^2<\frac{4\pi G\bar{\rho}}{\sigma_r^2}(0.45-\bar{Q}_M)
\end{equation}
with stability threshold

\begin{equation}
\bar{Q}_M=0.45.
\end{equation}
This corresponds to $Q_M=\bar{Q}_M/\sqrt{2}\approx0.3$ or, according to eq. (\ref{eq:QMQT}), roughly $Q_T\approx 1$, 
assuming $\alpha\sim1$ and $f_g\sim1$.  
Thus, 
the entire disk has a lower stability threshold than found specifically near the mid-plane, where $Q_M=1$ applies regardless of the vertical density distribution and regardless of the type of perturbation ($\S$~\ref{sec:3dmidplane}).  

As examined in the next section, the introduction of wave-like behavior ($k_z\neq0$) out to $z=\pm\infty$ modifies this threshold, but only significantly when $k_z>>k$ and the radial self-gravity force is weakened.  
Thresholds for perturbations that are finite (and do not extend to $\pm\infty$), on the other hand, tend to be raised when either the velocity dispersion is highly non-isotropic $\alpha<<1$ or the perturbation is present only well inside $h$ ($h_1/h<<1$).  

\subsection{Zero Vertical Mass Flux Infinite (Non-WKB) Wave Perturbations}\label{sec:infwave}
Now consider vertical wave perturbations that include phase variation ($k_z\neq 0$) 
but still fall off with height above the mid-plane, to satisfy the GLB 'no-mass flux at infinity' condition.  In this case, Poisson's equation implies % is modified.  That is, for these perturbations we will no longer require $k_z z_p>>1$.

\begin{equation}
\Phi_1=-\frac{4\pi G\rho_1}{k^2+k_z^2-T^2}
\end{equation}
with $T$ the same as in the previous section.  The potential is thus weakened with the introduction of non-zero $k_z$.  
This weakening is minimal in scenarios with $k>>k_z$, which are identical to the GLB scenario.  But when $k<<k_z$, the radial self-gravity term in the dispersion relation is considerably smaller.

After integration, the 2D dispersion relation becomes 

\begin{equation}
\bar{\omega}^2=
\begin{cases}
\kappa^2-4\pi G\rho_c\gamma_T \frac{k^2}{k_z^2}+\sigma_r^2k^2 &\text{$k<<k_z$}\\
\kappa^2- 4\pi G\rho_c\gamma_T +\sigma_r^2k^2  &\text{$k>>k_z$}
\end{cases}
\end{equation}
where %$\bar{\rho}=\rho_c/\sqrt{2}$ in the case of the Gaussian profile and 
$\gamma_T\approx 0.3$ (see previous section).  %or $\zeta=2/3$ in the case of the $\textrm{sech}^2(z)$ profile.  

These two regimes yield the stability condition ($\bar{\omega}^2<0$) that can be written as

\begin{equation}
k^2<\bar{k}_{J,r}(\zeta^2\sqrt{2}\gamma_T-\bar{Q}_M),
\end{equation}
assuming $k/k_z$ is a fixed ratio.  Here $\zeta=1$ ($k>>k_z$) or $\zeta=k/k_z$ ($k<<k_z$) and the (radial) Jeans wavenumber $\bar{k}_{J,r}=4\pi G\bar{\rho}/\sigma_r^2$. 

Rotation thus acts to stabilize above a threshold

\begin{equation}
\bar{Q}_M=
\begin{cases}
\gamma_T\sqrt{2}(k/k_z)^2& \text{$k<<k_z$}\\
\gamma_T\sqrt{2}& \text{$k>>k_z$}.
\end{cases}
\label{eq:waveQ}
\end{equation}
The behavior of $\mathscr{F}=(1/\gamma_T)$ also implies that disks are stable wherever $k_z h>>1$ (in the regime $k<<k_z$) or $kh>>1$ (in the regime $k>>k_z$).

Thus we see that the impact of the wave nature of the vertical perturbation is different in the two regimes.  In the first ($k<<k_z$) scenario, the condition for instability can also be written as

\begin{equation}
k^2<\frac{\kappa^2}{\frac{4\pi G\bar{\rho}\sqrt{2}\gamma_T}{k_z^2}-\sigma_r^2}
\end{equation}
which can be solved as long as

\begin{equation}
k_z^2<\bar{k}_J^2\alpha^2\sqrt{2}\gamma_T
\end{equation}
in terms of $\bar{k}_J=\bar{k}_{J,r}/\alpha^2$.  
Instability in the radial direction is thus unable to proceed without instability in the vertical direction.

In the opposite $k>>k_z$ scenario, instability is nearly insensitive to the vertical direction and possible as long as

\begin{equation}
k^2<\bar{k}_{J}^2\alpha^2(\sqrt{2}\gamma_T-\bar{Q}_M)\approx k_J^2\alpha^2(\gamma_T-Q_M)
\end{equation}
in terms of $Q_M=\kappa^2/(4\pi G\rho_c)$ and $k_J^2=4\pi G\rho_c/\sigma_z^2$.  
Stability is indeed identical to the case of the generic vertical (non-WKB) perturbation considered by GLB and here in $\S$ \ref{sec:infGLB}. 

It is notable that, in the long-wavelength regime $k<<k_z$, the $Q_M$ threshold in eq. (\ref{eq:waveQ}) is always lower than $\sqrt{2}\gamma_T\approx 0.45$, which makes it lower than the $Q_T$=1 threshold (see previous section).  
As examined in the next section, $Q_T$=1 can be viewed as the highest threshold that applies to extended perturbations in the limit of very small $h$ (or small $\sigma_z$ or highly non-isotropic velocity dispersions).  
Accounting for the 3D nature of the disk, the $Q_T$ threshold is lowered, as also indicated by the lowered $Q_M$ calculated in this section.  

\subsection{Finite WKB-wave Perturbations}\label{sec:finWKB}
To illustrate how the threshold $Q_M=1$ endemic to the mid-plane transforms smoothly in to the $Q_T$=1 threshold characteristic of the total destabilization of the disk, this section calculates the 2D dispersion relation for perturbations that extend a finite distance around the mid-plane.  
As discussed in $\S$~\ref{sec:perturbations}, truncations represent an opportunity to describe perturbations with amplitudes that vary more slowly than $\rho_0$ in the vertical direction (since they would drop to zero before the requirement $\rho_1/\rho_0<<1$ is violated).  This has the practical advantage that perturbations can be examined using the WKB approximation, and the amplitude variation can be arbitrarily with $z$ as long as it is slow, i.e. $k_z z_p>>1$.  
More critically, since $k_z>>T$, these perturbations entail higher self-gravity than infinite wave perturbations, enhancing the possibility for growth.  

However, unlike for the infinite perturbations with unrestricted $k$ and $k_z$, for finite perturbations the boundary condition couples  $k$ to $k_z$ and ties them both to the perturbation extent $h_1$.  This  puts a strong limit on $k_z$ in the short regime in particular, preventing $k$ from dropping below $1/h$ (when $h_1<<h$).  Thus we can expect perturbations in the short regime to be more readily stable than in the long regime, which is the reverse of the scenario predicted for infinite wave perturbations in $\S$~\ref{sec:infwave}.  

Finite perturbations also have the influential property that they entail mass flux through the perturbed region, as indeed, integration of the vertical terms even in the limit $h_1/h>>1$ does not necessarily yield zero.  
This is captured here by the 2D stability condition 

\begin{equation}
\kappa^2+\left(\frac{-4\pi G\rho_c F_r(x)}{k^2+k_z^2}+\sigma_r^2\right)k^2 +\left(\frac{-4\pi G\rho_cF_z(x)}{k^2+k_z^2}+\sigma_z^2\right)k_z^2 <0
\end{equation}
calculated by integrating the 3D dispersion relation over the vertical direction with bounds $\pm h_1$ and then identifying when $\omega^2<0$ (see $\S$ \ref{sec:3dmidplane}).  
Here $x$=$h_1/h$ and the factors $F_z(x)$ and $F_r(x)$ (see below) depend on the vertical density distribution of the unperturbed disk and the perturbation itself.  

For demonstration purposes (and for the sake of analytical simplicity), below focuses on the basic scenario in which the perturbation amplitude is constant.  This is a good approximation for any perturbations with amplitudes that vary more slowly with $z$ than $\rho_0$.  Indeed, for most other 'slow' choices, $F_r(x)$ and $F_z(x)$ recover essentially identical behavior in the limits $h_1/h<<1$ and $h_1/h>>1$ as determined in the constant amplitude case, even if their functional forms differ in detail from what is presented below.  

Combining this slow (constant) perturbation amplitude with the Gaussian vertical profile associated with the nominal weakly-self gravitating equilibrium disk, vertical integration yields 

\begin{equation}
F_z(x)=e^{-x^2/2}\label{eq:fzofx}
\end{equation}
and
\begin{eqnarray}
F_r(x)&=&e^{-\frac{h^2k_z^2}{2}}\frac{k_zh}{2\sin{k_zh_1}}\sqrt{\frac{\pi}{2}}\label{eq:frofx}\\
&\hspace{1cm}&\times\left[\textrm{Erf}\left(\frac{x-ik_zh}{\sqrt{2}}\right)+\textrm{Erf}\left(\frac{x+ik_zh}{\sqrt{2}}\right)\right]\nonumber\\
&\approx& \frac{1}{x}\sqrt{\frac{\pi}{2}}\textrm{Erf}\left(\frac{x}{\sqrt{2}}\right)\textrm{\hspace{0.5cm}$k_z h<<1$ and $k<<k_z$}\nonumber\\
&\approx& k_z h\sqrt{\frac{\pi}{2}}\textrm{Erf}\left(\frac{x}{\sqrt{2}}\right)\textrm{\hspace{0.28cm}$k_z h<<1$ and $k>>k_z$}\nonumber
\end{eqnarray}
using that $\sin{k_zh_1}\sim k_z h_1$ in the limit $k_z h_1<<1$ appropriate for our adopted finite perturbations in the regime $k<<k_z$ or $\sin{k_zh_1}=1$ when $k_z=\pi/(2h_1)$ as required for $k>>k_z$.  Note that, in the limit $k_zh>>1$, $F_r(x)\rightarrow 0$.

\begin{figure}[t]
%\begin{flushleft}
%\vspace*{-.15in}
\begin{center}
\begin{tabular}{c}
\includegraphics[width=0.915\linewidth]{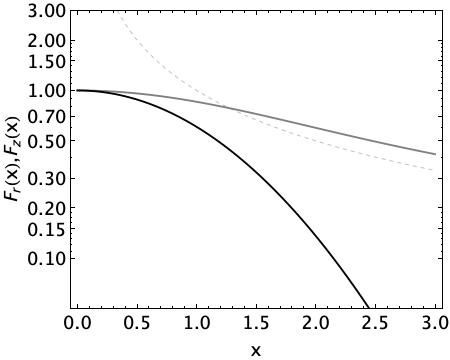}%{fzfrplot.pdf}%&\includegraphics[width=0.495\linewidth]{ratesfg2.pdf}
\end{tabular}
\end{center}
%\vspace*{-.15in}
\caption{Illustration of the behavior of $F_z(x)$ (black) and $F_r(x)$ (gray) as a function of $x=h_1/h$ for finite wave perturbations in 3D flattenend rotating disks.  These functions appear in the 2D dispersion relation for finite vertical WKB perturbations with vertical extents $\pm h_1$ and amplitudes that vary slowly as a function of height $z$ from the mid-plane ($1/z_p<1/z_d$); see $\S$ \ref{sec:finWKB}.   This example adopts the functional forms for $F_z(x)$ and $F_r(x)$ given by eqs. (\ref{eq:fzofx}) and (\ref{eq:frofx}), determined in the case of a constant amplitude perturbation.  A third dashed gray reference lines shows $1/x$.  }
\label{fig:fzfr}
%\end{flushleft}
\end{figure}

These factors simplify considerably in the limits $h_1/h<<1$ or $h_1/h>>1$, becoming
%$h_1\rightarrow 0$ or $h_1\rightarrow \infty$.  

\begin{eqnarray}
F_z(x)&\approx&
\begin{cases}
1 & \text{$x<<1$}\\%$x\rightarrow 0$}\\
0& \text{$x>>1$}\\%$x\rightarrow \infty$}
\end{cases}\\
F_r(x)&\approx&
\begin{cases}
\begin{cases}
1 & \text{$k<<k_z$}\\%$x\rightarrow 0$}\\
\frac{\pi}{2} & \text{$k>>k_z$}\end{cases}&\text{$x<<1$}\\
\begin{cases}
\sqrt{\pi/2}(1/x)& \text{$k<<k_z$}\\
\sqrt{\pi/2}k_z h& \text{$k>>k_z$}%$x\rightarrow \infty$}
\end{cases}&\text{$x>>1$}
\end{cases}
\end{eqnarray}

The behavior of $F_r(x)$ and $F_z(x)$ in these opposite limits is key to the recovery of both stability above $Q_M$=1 when $h_1/h<<1$ and stability above $Q_T=1$ when $h_1/h>>1$.  

\subsubsection{$h_1/h>>1$: The Lin-Shu Dispersion Relation and $Q_T$=1 Threshold}\label{sec:largeh1}
In the limit $x>>1$, the 2D 
stability condition reads 
 \begin{equation}
0>\kappa^2-\frac{4\pi G\rho_c k^2}{k^2+k_z^2}\sqrt{\frac{\pi}{2}}\frac{h}{h_1}+\sigma_r^2k^2+\sigma_z^2k_z^2,
\end{equation}
which can be used to predict the behavior of stability in the regimes $k<<k_z$ and $k>>k_z$.  \\

\noindent\underline{The Long Wavelength Regime $k<<k_z$}\\
\vspace*{-.1in}

In the first case, inserting the specific relation between $k$ and $k_z$ appropriate for this scenario ($k=k_z^2h_1$), the factor 
\begin{equation}
\frac{k^2}{(k^2+k_z^2)}\approx \frac{k^2/k_z^2}{(k^2/k_z^2+1)}\approx \frac{kh_1}{1+kh_1}
\end{equation}
which can be approximated by $kh_1$ to lowest order. (Note that in this regime, $k_zh_1<<1$ and $kh_1<<1$.)  
The 2D dispersion relation thus yields the instability condition 

\begin{equation}
0>\kappa^2-2\pi G\Sigma_c k+\sigma_r^2k^2+\sigma_z^2\frac{k}{h_1}\label{eq:linshu}
\end{equation}
where $\Sigma_c=\rho_c\sqrt{2\pi}h$ for our unperturbed disk.  

In the limit $h_1>>h$ (or in the limit $\sigma_z\rightarrow 0$), the fourth term above

\begin{equation}
\sigma_z^2\frac{k}{h_1}=\frac{\sqrt{8\pi} G\Sigma}{f_g} \left(\frac{h}{h_1}\right)k
\end{equation} 
and can be neglected, and the 2D dispersion relation is the axisymmetric Lin-Shu dispersion relation, which can solved under the condition $\omega^2$$<$0 for the onset of instability to obtain the familiar requirement 

\begin{equation}
k<\frac{k_T}{2}\left[1\pm (1-Q_T^2)^{1/2}\right]\label{eq:ktoomre}
\end{equation} 
for  unstable (growing) modes, in terms of 

\begin{equation}
Q_T=\frac{\sigma_r\kappa}{\pi G\Sigma}
\end{equation}
and the wavenumber associated with the Toomre length 

\begin{equation}
k_T=\frac{2\pi G\Sigma}{\sigma_r^2}. 
\end{equation} 
The inequality in eq. (\ref{eq:ktoomre}) gives us the well-known stability condition $Q_T>1$ that describes the suppression of long-wavelength instabilities by rotation.  

The threshold for stability is lowered when the disk is allowed to have some thickness (or non-negligible $\sigma_z$), weakening the perturbed gravitational force in the plane \citep[e.g.][]{toomre, jogsolomon,ghoshjog}.  
This is evident here by letting $\alpha>0$, or keeping all terms to lowest order in $h/h_1$, such that the condition for instability becomes

\begin{equation}
0>\kappa^2-\left(2\pi G\Sigma_c-\frac{\sigma_z^2}{h_1}\right) k +\sigma_r^2k^2
\end{equation}
with the term in parentheses corresponding to weakened self-gravity.  
This can be solved to yield

\begin{equation}
k<\left[\frac{k_T}{2}-\frac{\alpha^2}{h_1}\right](1\pm \sqrt{1-Q_{T,t}})\label{eq:klongiso}
\end{equation}
in terms of the thickened $Q$ parameter % $Q_{T,t}$ 

\begin{equation}
Q_{T,t}=\frac{Q_T^2}{\left(1-\frac{\alpha^2}{k_Th_1}\right)^2}.
\end{equation}
In the limit $\alpha<<(k_T h_1)^{1/2}$, eq. (\ref{eq:klongiso}) implies that rotation suppresses the growth of 3D perturbations above a threshold

\begin{equation}
Q_T=\left(1-\frac{\alpha^2}{k_Th_1}\right)
\end{equation}
in terms of the velocity anisotropy parameter $\alpha=\sigma_z/\sigma_r$. This is approximately $Q_T=1-h/h_1$, since $k_T\approx\sqrt{\pi/2}/h$.  Stability for a 3D disk with nearly isotropic velocity dispersion is thus predicted to set in above a threshold that is slightly lower than $Q_T=1$.  

Note that this constitutes a higher threshold than calculated for infinite perturbations in the same regime, for which $Q_T\approx 2(\gamma_T)^{1/2}(k/k_z)\approx(k/k_z)$.  This reflects the stronger self-gravity associated with the slowly varying amplitudes in the present case compared with fall-off required in the infinite case.  

In the opposite limit  $\alpha>>k_T h_1$, instability (which eq. [\ref{eq:klongiso}] implies would require $Q_{T,t}<0$) is entirely suppressed since $Q_{T,t}$ is positive definite. \\

\noindent\underline{The Short Wavelength Regime $k>>k_z$}\\
\vspace*{-.1in}

In the opposite regime where $k>>k_z=\pi/(2h_1)$, the 2D dispersion relation implies that instability can occur as long as
 
\begin{equation}
0>\kappa^2-\frac{4\pi G\rho_c k^2}{k^2+k_z^2}\sqrt{\frac{\pi}{2}}k_z h+\sigma_r^2k^2+\sigma_z^2k_z^2
\end{equation}
or approximately
 \begin{equation}
0>\kappa^2-2\pi G\Sigma_c k_z+\sigma_r^2k^2+\sigma_z^2k_z^2\label{eq:2Dshortbigh1}
\end{equation}
to lowest order in $k_z/k$.  

This can be treated as a condition on $k_z$ (and $h_1$) 
in a manner that parallels the condition for Toomre instability, i.e. 

\begin{equation}
k_z<\frac{k_T}{2}(1\pm \sqrt{1-Q_{T,ep}})\label{eq:klongiso}
\end{equation}
in terms of  % $Q_{T,t}$ 

\begin{equation}
Q_{T,ep}=Q_{T,z}^2\left(1-\frac{\sigma_r^2 k^2}{\kappa^2}\right)
\end{equation}
with $Q_{T,z}=Q_T(\sigma_z/\sigma_r)$.  
This suggests the stability threshold 

\begin{equation}
Q_{T,z}=\frac{1}{1+(R_{ep}k)^2}
\end{equation}
in terms of the epicyclic radius $R_{ep}=\sigma_r/\kappa$.  When perturbations satisfy $R_{ep}k<<1$, the valid stability threshold remains at $Q_{T,z}\approx Q_T=1$.  In the limit $h_1>>h$, $k$ in this regime will indeed always be smaller $1/h$, which can be expected to be near $1/R_{ep}$.  

\subsubsection{$h_1/h<<1$: The Mid-plane Dispersion Relation and $Q_M$=1 Threshold}\label{sec:smallh1}
In the limit $h_1/h<<1$, it is perhaps not surprising that % as$x\rightarrow 0$, 
the 2D dispersion relation is identical to the dispersion relation very near the mid-plane calculated in $\S$~\ref{sec:3dmidplane}. \\

\noindent\underline{The Long Wavelength Regime $k<<k_z$}\\
\vspace*{-.1in}

In the limit $k<<k_z$ and $h_1<<h$, instability is found to require 

\begin{equation}
0>\kappa^2-4\pi G\rho_c +\sigma_r^2 k^2+\sigma_z^2k_z^2\label{eq:2dmid}.
\end{equation}
adopting $F_r(x)$ and $F_r(x)$ appropriate for the present scenario.
This suggests the general condition 

\begin{equation}
k_S^2=k^2+k_z^2<k_J^2(1-Q_M)
\end{equation}
when the velocity dispersion is isotropic, and a stability threshold $Q_M=1$.

More precisely, eq. (\ref{eq:2dmid}) is quadratic in $k$ and yields the condition

\begin{equation}
k_z^2h_1\approx k<-\frac{\alpha^2}{2h_1}\left(1\pm\sqrt{1-Q_{2D,mid}}\right)
\end{equation}
where the parameter

\begin{equation}
Q_{2D,mid}=\left(\frac{2f_gh_1}{\alpha h}\right)^2(Q_M-1).
\end{equation}
For this to yield a real solution for $k_z$, $Q_{2D,mid}>0$, once again yielding the stability threshold $Q_M=1$. \\

\noindent\underline{The Short Wavelength Regime $k>>k_z$}\\
\vspace*{-.1in}

In the short limit $k>>k_z=\pi/(2h_1)$, 

\begin{equation}
0>\kappa^2-\frac{4\pi G\rho_c k^2}{k^2+k_z^2}\left(\frac{\pi}{2}\right)+\sigma_r^2 k^2+\sigma_z^2k_z^2-\frac{4\pi G\rho_c k_z^2}{k^2+k_z^2}\label{eq:2dmid}.
\end{equation}
which is approximately

\begin{equation}
0>\kappa^2-4\pi G\rho_c \frac{\pi}{2} +\sigma_r^2 k^2+\sigma_z^2k_z^2\label{eq:2dmid}.
\end{equation}
 to lowest order in $k_z/k$.

This yields the following condition on $k_z$ (or $h_1$)

\begin{equation}
k_z^2<k_J^2\left(\frac{\pi}{2}-Q_M-Q_M\left(R_{ep}k\right)^2\right)
\end{equation}
suggesting the stablity threshold

\begin{equation}
Q_M=\frac{\frac{\pi}{2}}{1+\left(R_{ep}k\right)^2}.
\end{equation}
Since $k$ is not guaranteed to be smaller than $1/h$ (or $R_{ep}$) in the limit $h_1<<h$, however, perturbations are easily stabilized, with a stability threshold that is considerably lower than $Q_M=1$.  

\subsection{Summary}
In the previous sections the stability threshold for 3D rotating disks (above which disks are stabilized) was found to be influenced by the presence of a vertical perturbation: whether it has wave- or non-wave traits, how strongly the amplitude varies, how far it extends in the vertical direction, and its relation to the scale height of the unperturbed disk.  {Lower predictions for the threshold are a mark of stability (since the threshold is more easily surpassed), while higher thresholds suggest that the disk is more unstable. The reference adopted for this study is the overall threshold $\kappa^2/\pi G\bar{\rho}=0.45$ (near $Q_T=1$) determined by GLB for stability out to $z=\pm\infty$ in the presence of a non-wave infinite vertical perturbation that falls off with $z$ like the unperturbed disk density.  

For these infinite perturbations, adding phase variation (wave-like behavior) as a rule reduces the self-gravity of the perturbation and thus lowers the stability threshold (signifying a more easily stabilized disk), although this is a negligible change when $k>>k_z$.  The self-gravity can be increased again (even with wave-like behavior) when the perturbation has a more slowly varying amplitude than infinite perturbations and is also necessarily finite (so as to avoid violating the requirement $\rho_1/\rho_0<<1$).  In this manner, finite but extended ($h_1/h>>1$) long-wavelength WKB perturbations are shown to have a higher threshold than when they are infinite, with more rapidly varying amplitudes.  The The threshold in this case is exactly $Q_T=1$, signifying an increase back up near to the level $\kappa^2/\pi G\bar{\rho}=0.45$.  As ever, though, allowing for non-negligible thickness lowers this threshold (see $\S$ \ref{sec:largeh1}).    

An even more consequential factor, capable of shifting the stability threshold {\it above} $Q_T\sim1$ (or $\kappa^2/\pi G\bar{\rho}=0.45$) -- and widening the avenue for instability -- is the extent of the perturbation and its relation to the scale height of the unperturbed disk.   This is a consequence of the sensitivity of vertical stability to height above the mid-plane, which was identified in the case of generic wave or non-wave perturbations in $\S$ \ref{sec:3dmidplane} using the 3D dispersion relation.  As a rule, perturbations near the mid-plane or finite (WKB) perturbations extending only out to $h_1<<h$, are subject to a stability threshold $Q_M=1$ which corresponds to $Q_T=2/(\alpha f_g^{1/2})\approx 2$.  As the perturbation vertically extends across more of the disk, its stability threshold is lowered back to $Q_T\sim1$.

In this light, disks are expected to be more stable to features that pervade the entire vertical extent of the disk than to perturbations local to the mid-plane.  In other words, it is harder to prevent fragmentation near the mid-plane at a given $Q_T$ than it is to stop the whole disk from becoming unstable.  

From this perspective, there are two stability regimes of consequence  for the appearance of disks. These are referred to in what follows as either `partial 3D', in which the radial instability is localized around the mid-plane (but still limited to scales larger than the Jeans length), subject to threshold $Q_M=1$, or `total 2D', in which radial instability is present throughout the entire vertical extent of the disk and the relevant threshold is $Q_T$=1.  The latter choice is meant to bring to mind that $Q_T=1$ is the threshold calculated for a 2D disk with perturbation restricted to the plane.  

\section{The onset of partial 3D vs. total 2D instability}\label{sec:3Dv2D}
\begin{figure*}[t]
%\begin{flushleft}
%\vspace*{-.15in}
\begin{center}
\begin{tabular}{cc}
\includegraphics[width=0.495\linewidth]{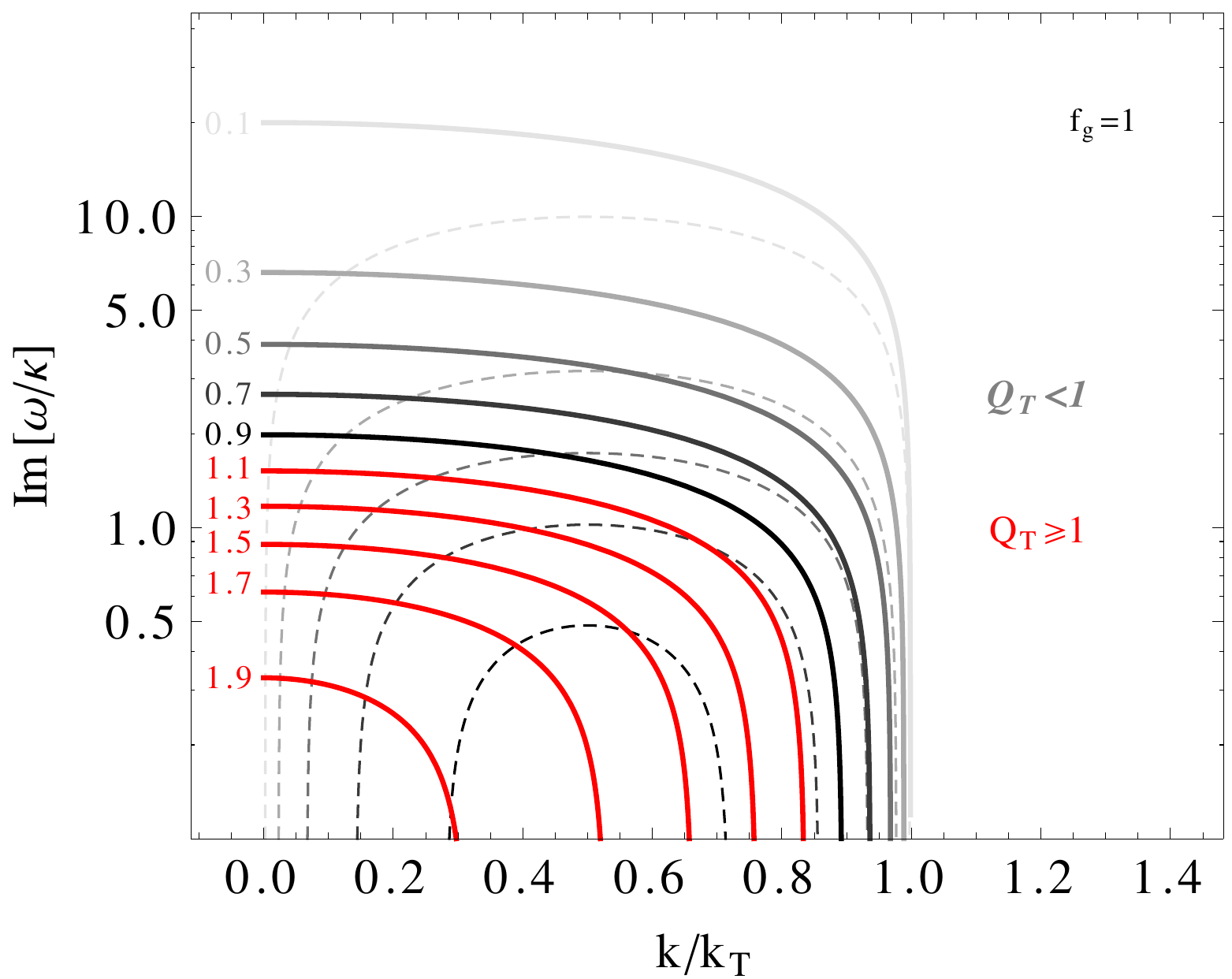}&\includegraphics[width=0.495\linewidth]{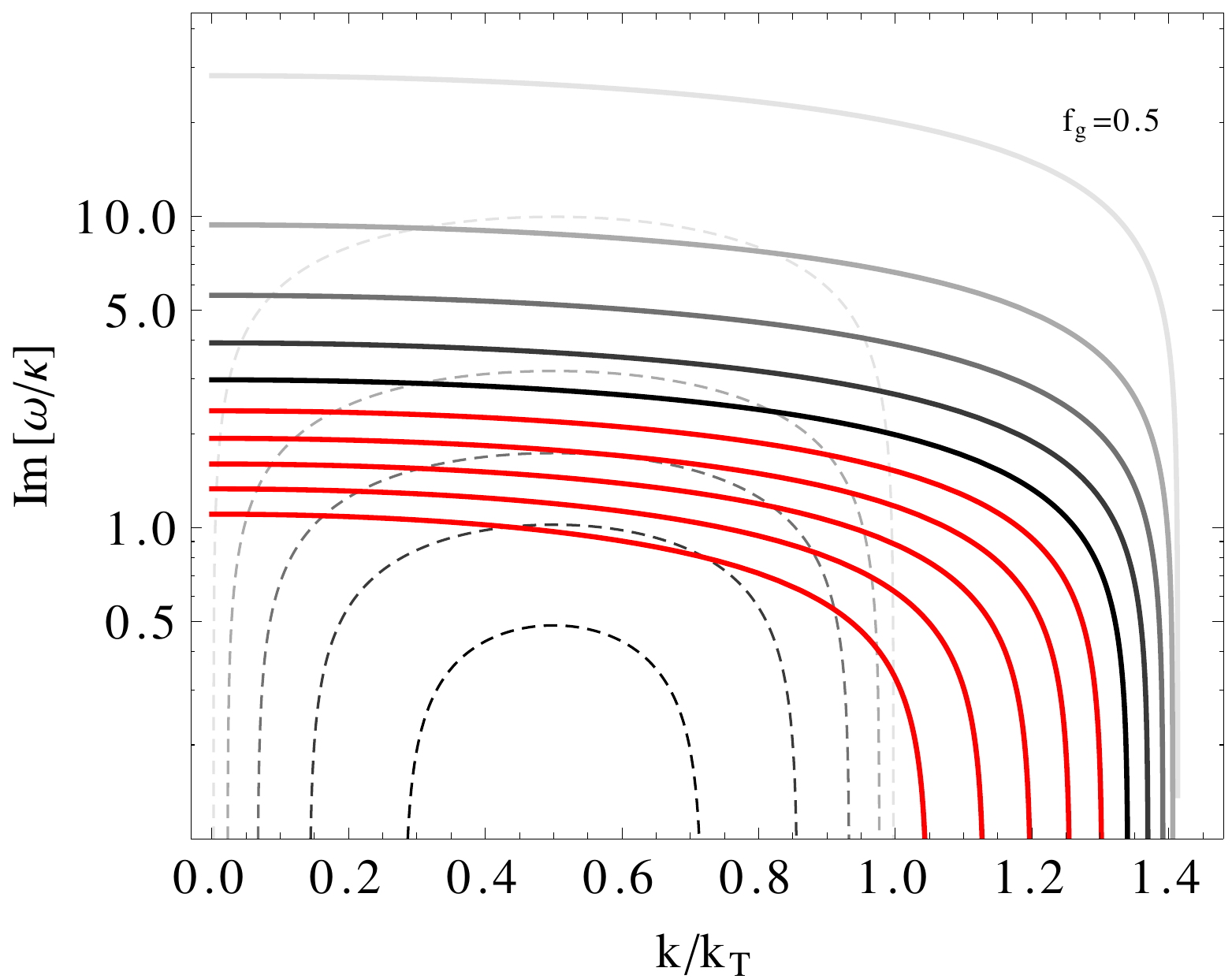}
\end{tabular}
\end{center}
%\vspace*{-.15in}
\caption{Growth rates for 2D Toomre instabilities (dashed lines) and for 3D instabilities at the mid-plane (solid lines) in a fully self-gravitating flattened disk ($f_g$=1; left) and a weakly self-gravitating flattened disk ($f_g$=0.5; right).  The gray scaling of both sets of curves increases with increasing $Q_T$ (from 0 to 1). Curves with $Q_T\geq$1 are shown in red.  As $f_g$ goes down, instabilities undergo faster growth on smaller scales at fixed $Q_T$, making 3D fragmentation more prominent than 2D structures in weakly self-gravitating disks.  }
\label{fig:rates}
%\end{flushleft}
\end{figure*}
\subsection{Overview}\label{sec:longshortsummary}
In the previous sections the 2D and 3D dispersion relations were used to show that there are two relevant thresholds for describing 3D disk instability.  
The first threshold -- the Toomre threshold 

\begin{equation}
Q_T\equiv \frac{\sigma_r\kappa}{\pi G\Sigma}=1\nonumber
\end{equation}
(see $\S$~\ref{sec:largeh1}) -- applies to the disk's total ability to destabilize, across the entire disk out to $z\rightarrow\pm\infty$.  The second threshold 

\begin{equation}
Q_M\equiv\frac{\kappa^2}{4\pi G\rho_c}=1\nonumber
\end{equation}
(see $\S\S$~\ref{sec:3dmidplane} and $\S$\ref{sec:smallh1}) applies to the 3D instability at the mid-plane, which more closely resembles Jeans instability than Toomre instability.  

Under most normal circumstances \citep[adopting the typical masses and rotational properties of disk galaxies; see e.g. ][]{meidt18} the $Q_M$ threshold that applies in 3D is higher than the 2D Toomre $Q_T$ threshold.  The two additional degrees of freedom introduced by the vertical direction more than compensate for the stabilizing influence of disk thickness, similar to the role that a secondary (stellar) disk has been shown to play on gas stability \citep[e.g.][]{ko07}.  The difference in 2D and 3D thresholds signifies that 
fragmentation at the mid-plane should be possible even where the Toomre threshold is surpassed.     

Turbulent dissipation and cooling also favor gravitational instability even where gas is Toomre stable \citep{gammie01,elm11}, i.e. once the gas velocity dispersion (and pressure support) is lowered through turbulent dissipation. The present work shows that (even without incorporating dissipation or cooling), pressure forces can be overcome by self-gravity preferentially at the disk mid-plane, where the gas density is approximately constant in equilibrium.  

Considering exclusively the basic equilibrium scenario discussed in this work (ignoring cooling, turbulence dissipation and magnetic forces), whether disk fragmentation is ultimately a partial 3D or total 2D process 
can be expressed in terms of the growth rates of perturbations and proximity to the critical density associated with each stability threshold, as discussed below.  

Before proceeding, it is worth noting that the onset of instability triggered by 3D perturbations is unaffected by a (vertical) rotational lag in either the partial or total instability regimes. 
Non-axisymmetry also does not alter the $Q_M$=1 threshold (Appendix \ref{sec:largeh1}), although it has been shown to modestly increase the $Q_T$ threshold \citep[][Appendix \ref{sec:largeh1}]{laubertin, bertinplus, grivgedalin}).  The increase is, however, substantially smaller than the increase in the $Q_T$ threshold represented by $Q_M=1$.  
Indeed, $Q_T$=1 is expected to remain valid in most scenarios with $m>0$ \citep{BT}.   

\subsection{The Critical Density}\label{sec:rhocrit}
The molecular gas disks of nearby galaxies are observed to sit near $Q_T\sim$2 \citep{leroy08}, placing them just at the threshold for stability at the mid-plane, i.e. $Q_M=1$.  In general, proximity to $Q_M$=1 depends on the degree of self-gravitation; the more weakly self-gravitating the disk, the quicker the $Q_M$=1 threshold is passed.  
Again letting 
%\begin{equation}
$f_g=\rho_0/(\rho_0+\rho_b)$ in terms of the background density $\rho_b$, 
%\end{equation}
then $Q_M=\kappa^2/(4\pi G\rho_c)$ can be rewritten as 

\begin{equation}
Q_M=\frac{1}{f_g}\left(\frac{\kappa}{\nu}\right)^2
%&=&f_g\frac{Q^2}{4}.
\end{equation}
where $\nu=4\pi G(\rho_c+\rho_b)$.  
Considering that typically $\kappa/\nu\sim0.5$ in nearby star forming (disk) galaxies, then wherever the gas fraction $f_g$ is below $\sim$0.25, $Q_M$ will exceed unity.  Thus a background potential can be viewed as a source of stability for any embedded disk, suppressing structures unless the disk's density is increased above a critical value 

\begin{equation}
\rho_{crit}=\kappa^2/(4\pi G).  
\end{equation}
This has also been discussed by \cite{jog2014}, who emphasized that in weakly self-gravitating disks, rotation and the epicyclic frequency $\kappa$ are decoupled from the embedded disks's mass distribution (tracking instead the dominant background distribution).  This necessitates a comparable increase in the local disk density for instability to occur.  

In many disks, $\rho_{crit}$ is a lower threshold to pass than the volume density associated with the Toomre critical density
$\Sigma_{crit,T}=\sigma_r\kappa/(\pi G)$ since

\begin{equation}
\rho_{crit}=\rho_{crit,T}\frac{f_g\alpha}{4}  
\end{equation}
where $\rho_{crit,T}=\Sigma_{crit,T}/(2h)$.  This makes disk instability easier near the mid-plane than overall.  

\subsection{Growth Rates}\label{sec:growthrates}
The closer molecular disks are to the critical density, the lower $Q_M$ and the more favorable they are to small-scale Jeans-like 3D instabilities.  The prominence of the structures that result from this gravitational instability can be assessed by considering the growth rates of different perturbations over different scales, under a given set of conditions.  

For 3D perturbations endemic to the mid-plane, growth rates can be approximated as

\begin{equation}
\omega_{3D}^2/\kappa^2\approx
1-\frac{4}{Q_T^2}\frac{1}{f_g}+\frac{4}{Q_T^2}\left(\frac{k}{k_T}\right)^2+\frac{4}{Q_T^2}\left(\frac{k_z}{k_T}\right)^2 \label{eq:growth3d}
\end{equation} 
taking $\omega_{min}^2$ as the lower bound on $\omega^2$ and rewriting $Q_M$ in terms of $Q_T$. Here $\alpha$ is set to unity and $m=0$ is adopted for simplicity.  The maximum growth rates at a given $k$ are associated with the largest vertical perturbations $k_z<<k_T$, as exclusively considered below.   

Under the same conditions ($m=0$, $\alpha=1$ and using that $Q_M=\alpha^2Q_T^2 f_g/4$), the growth rates of 2D perturbations predicted by the Lin-Shu dispersion relation can be written as

\begin{equation}
\omega_{2D}^2/\kappa^2\approx
1-\frac{4}{Q_T^2}\left(\frac{k}{k_T}\right)+\frac{4}{Q_T^2}\left(\frac{k}{k_T}\right)^2\label{eq:growth2d}.
\end{equation} 

Note that neither this expression or eq. (\ref{eq:growth3d}) is expected to be valid when $k$ is small, since both are derived assuming 
$kR>>1$ (Binney \& Tremaine 2008).  These expressions also only apply 
to the fastest growing perturbations with negligible vertical rotational lag (see Appendix \ref{sec:rotationallagappendix}). %
The growth rates of tightly-wound non-axisymmetric $m\neq0$ instabilities (estimated in Appendix \ref{sec:largeh1}) are similar.     

Figure \ref{fig:rates} compares the growth rates Re($i\omega$) of instabilities on different scales in the partial 3D and total 2D regimes. 
Perturbations have mostly comparable growth rates in the two regimes over the entire range in $k$. But there is a scenario signified by $Q_T>1$ in which only 3D perturbations at the mid-plane can grow, demarcated by the red curves.  Like the growth in the 2D regime, 3D growth appears everywhere above the Jeans scale.  But, under certain conditions, this growth can occur below $k_T$, as illustrated in the right panel of the figure.   These are situations where the disk is only weakly self-gravitating ($f_g<1$), and the Jeans length exceeds the scale height (since $\lambda_J=h/f_g$). In these cases, 3D instabilities are still able to occur on small scales, closer to $h$ and $\lambda_J$ than $\lambda_T$.  Embedding the gas disk in a dominant external potential therefore does not suppress small scale structure.  It may even favor vertical perturbations of the kind adopted in this work.  
 
Another characteristic of 3D mid-plane instability in the low-$f_g$ scenario is faster growth at fixed $Q_T$ than when $f_g$=1.  This can make weakly self-gravitating disks more prone to 3D mid-plane instabilities than 2D instabilities.  Since $Q_T$ increases as $f_g$ decreases (and equilibrium velocity dispersions reflect more and more the background potential), a given $Q_T$ corresponds to a lower $Q_M$ as $f_g$ decreases.  The result is faster growth on smaller scales.  

\subsection{Discussion}
\subsubsection{Molecular Clouds as Instabilities}
The modified criterion presented here applies to axisymmetric ($m$=0) ring instabilities and non-axisymmetric instabilities (Appendix \ref{sec:largeh1}).  It should thus provide a useful diagnostic for the development of the rich small scale structure observed in gas disks, in much the same way that the axisymmetric Toomre criterion serves as a gauge of stability in general, including to non-axisymmetric stability.  

Indeed, following \cite{wangsilk94}, the growth rates of 3D instabilities in Figure \ref{fig:rates} provide an estimate for the cloud formation rate.  Given the properties of molecular disks and stellar disks in nearby galaxies, $f_g\approx0.5$ \citep[see e.g.][]{sun20,meidt21} and eq. (\ref{eq:growth3d}) predicts that clouds and cloud complexes can form rapidly, at a rate $\sim2\kappa$ or with a characteristic formation timescale of $\sim t_{orb}/3$.  

A prerequisite for the growth of any cloud structures is still the availability of vertical seed perturbations.  Gas disks embedded in thicker gas and stellar disks would seem to readily encounter such perturbations, which might take the form of stellar bar and spiral arms or stellar overdensities, in general (including stellar clusters), phase transitions, and pockets of gas that participate in the disk-halo flow or respond to triggers originating internal or external to the disk.  The multi-scale impact of feedback from star formation can also be envisioned as prompting perturbations at or near the mid-plane and beyond \citep[e.g.][]{kimtigress}.  

\subsubsection{Instabilities in Numerical Simulations}
In principle, existing realistic multi-phase 3D numerical disk simulations (as opposed to razor-thin models) should already capture the 3D disk instability described in this work, although it may be easiest to recognize in the absence of, e.g., a fixed spiral pattern and when controlling for magnetic forces (not included in the present calculation).  These are important factors for cloud formation via collisions, the wiggle instability and magneto-Jeans instability instability, for example \citep{elm87,WadaKoda,ko06,dobbs08}.  

Cloud formation through gravitational instabilities assisted by turbulence dissipation and/or cooling \citep[e.g.][]{gammie96,elm11} is also in principle recoverable in modern 3D numerical simulations, but it may be distinguishable from 3D disk fragmentation as it would not necessarily favor regulation to a particular $Q_T$ value.  From the perspective adopted in this work, the more profound consequence of the turbulent nature of molecular gas is to allow the deep interiors of the pressure-supported cloud fragments formed via 3D instability to collapse into the dense cores that go on to form stars, as proposed by \cite{KM05} \citep[see also][]{padoan,fk12}.

\subsubsection{Instabilities in Stellar Disks}
To the extent that the dynamics of stellar disks can be represented by fluid mechanics \citep[e.g.][]{jogsolomon}, the modified stability criterion derived here has implications for their stability and structure as well.  A source of 3D perturbations with $k_zh\lesssim1$ may be less obvious than in the case of molecular gas disks, though (except in exceptional cases, like interactions), so even if locally $Q_M<1$, fragmentation near the disk scale height is not guaranteed.  Still, it may be interesting that the stellar component of nearby galaxies has been measured to have $Q_T\gtrsim2$ \citep{bottema,kregel,westfall14} and some numerical simulations suggest that fragmentation is suppressed only once similar values are reached \citep[see][and references therein]{grivgedalin}.  

In this context, it is notable that for self-gravitating systems, the $Q_M$ stability threshold is equivalent to a constraint on geometry.  The epicyclic frequency can be written as $\kappa^2\approx2V_c^2/R^2=(2/3)4\pi G\rho_{sphere}$ in terms of the circular velocity $V_c$ and the volume density $\rho_{sphere}$ that would be equivalent to arranging all the mass internal to $R$ in a sphere.  The flatter the arrangement, the lower $Q_M=(2/3)\rho_{sphere}/\rho_0$, thus indicating a preference for instability and fragmentation in flatter, disk-like geometries.

\section{Summary and Conclusions}
This paper examines the stability of disks to a diversity of 3D perturbations, with the aim of describing situations apart from the Lin-Shu density wave scenario in which the waves are confined to an infinitely thin disk.  The chosen perturbations are meant to roughly represent the impact of events and processes taking place within gas disks as a consequence of their thickness and the fact that they are themselves i/ embedded within more extended gas and stellar disks and ii/ subject to on-going events like phase transitions and feedback from star formation.  

For the equilibrium disks under consideration (wherein pressure and gravity are the two most important factors, neglecting cooling, dissipation and magnetic forces), the inclusion of a vertical perturbation is found to be consequential.  This is fully characterized using the 3D dispersion relation ($\S$~\ref{sec:3ddispersionRelation}), which is shown to encode variations in disk stability with height above the mid-plane ($\S$~\ref{sec:verticalonly}).  This applies regardless of the chosen vertical form of the perturbation: with or without periodic (wave) components and either extending to infinity (as treated by GLB) or to a finite height above the mid-plane (and treatable with the WKB approximation).  

Near the mid-plane, in particular, where the unperturbed gas density is overall roughly constant, instability is found to proceed in a manner that is more Jeans-like than Toomre-like.  The onset of instability in this scenario is restricted to scales larger than the effective Jeans length (in the presence of thermal and non-thermal motions) in both the radial and vertical directions.  The instability is moreover subject to a modified threshold $Q_M=\kappa^2/(4\pi G\rho_c)=1$, or roughly $Q_T=2$, in terms of the Toomre $Q_T$, the radial epicyclic frequency $\kappa$ and the gas volume density $\rho_c$ at $z=0$.  This applies in the presence of a rotational lag (Appendix \ref{sec:rotationallagappendix}) or non-axisymmetry in the plane (Appendix \ref{sec:nonaxisym}). 

At locations well beyond a disk scale height $h$, however, the 3D dispersion relation describes characteristic stability.  This leads the total disk to be stabilized at a lower overall threshold than found endemic to the mid-plane.  The lowered threshold is, namely, the threshold obtained from the 2D dispersion relation ($\S$~\ref{sec:2dstability}), which is either $\kappa^2/(4\pi G\rho_c)\approx0.3$, as determined by GLB in the case of infinitely extended non-periodic vertical perturbations ($\S$~\ref{sec:infGLB}), or the Toomre threshold $Q_T=1$ obtained in this work using coupled radial and vertical wave perturbations in the limit of negligible disk scale height $h$ ($\S$~\ref{sec:finWKB}).   

The difference in the thresholds for partial and total 3D instability indicate that disks may be able to fragment at their mid-planes (above the Jeans length) even where the Toomre threshold is surpassed, as long as $Q_M<1$.  
The instabilities that are seeded at the mid-plane grow rapidly, comparable to Toomre instabilities, and 
with characterstic scales near the disk scale height in most scenarios of interest.  If we equate the formed fragments with molecular clouds stabilized from within by gas pressure, their formation is predicted to be fast, with a rate of approximately 2$\kappa$ and thus a characteristic  timescale of roughly $t_{orb}/3$ (given the properties of nearby galaxy disks).  This would make cloud formation compatible with fast destruction by early stellar feedback \citep{elm11,maclow17}.

Overall, considering the possibility of a broad variety of perturbations, the results of this study imply that pervasive gravitational instability is a characteristic of gas disks \citep[see also][]{elm11}, responsible for their rich multi-scale structure, the efficient conversion of ordered motion into turbulent motion and ultimately star formation.\\

Many thanks to the referee for a constructive, detailed review of the paper.  Thanks also to Arjen van der Wel and the members of the PHANGS (http://phangs.org) `Large-scale Dynamics Processes' Science Working Group for their feedback.   
\appendix
\section{The impact of a rotational lag on the conditions for instability}\label{sec:rotationallagappendix}
The main text exclusively considers perturbations for which the effect of a rotational lag is negligible.  Here precise bounds on the perturbations that meet this criterion are determined at the mid-plane from the 3D dispersion relation and overall from the 2D dispersion relation.  

For this calculation, the rotational lag in the perturbed radial velocity, which is proportional to 

\begin{equation}
\frac{dV_c}{dz}=-\frac{1}{2\Omega}\frac{d}{dz}\frac{d\Phi}{dr}
\end{equation} 
(since $d V_c^2/dz=2V_c dVc/dz=d(-Rd\Phi/dR)/dz$), is retained and evaluated assuming that the potential is a separable function of $z$ and $R$.  In this case, assuming that the disk is weakly self-gravitating and embedded in a background distribution with approximately constant density $\rho_b$, then 

\begin{equation}
\frac{d}{dz}\frac{d\Phi_0}{dR}=z\frac{d \nu^2}{dR}\equiv z\frac{\nu^2}{R_\nu}
\end{equation}
where $d\Phi_0/dz=\nu^2 z$, $\nu^2=4\pi G\rho_{b}$ and $R_\nu$ is defined as the scale length of the variation in $\nu^2$ with radius.  
\subsection{At the Mid-plane}
Consider a perturbation that is WKB-like at the mid-plane.  
With the rotational lag term included, the continuity equation becomes

\begin{eqnarray}
0&=&(-\omega+m\Omega)\rho_1+\frac{(-\omega+m\Omega)}{\Delta}C_r\nonumber\\
%&-&\frac{1}{\Delta(-\omega+m\Omega)}kk_z\left(\frac{-4\pi G\rho_1\rho_0}{k^2+k_z^2}+\sigma_z\right)\frac{d}{dz}\frac{d\Phi_0}{dR}\nonumber\\
&+&\frac{\nu^2L_z}{\Delta(-\omega+m\Omega)}\\
&+&\frac{C_z}{(-\omega+m\Omega)}
\end{eqnarray}
substituting in the expression for $v_{z,1}$ and
setting 

\begin{equation}
C_r=\left(\frac{-4\pi G\rho_1\rho_0}{k^2+k_z^2}+\rho_1\sigma_z\right)k^2
\end{equation}
and
\begin{equation}
L_z=-kk_z\frac{z}{R_p}\left(\frac{-4\pi G\rho_1\rho_0}{k^2+k_z^2}+\rho_1\sigma_z\right).\label{eq:lz}
\end{equation}

The 3D dispersion relation is again quadratic in $\omega^2$ (as in $\S$ \ref{sec:3dmidplane}), now with solution

\begin{equation}
\omega^2=\frac{\omega_{min}^2}{2}\left(1\pm\sqrt{1+4\frac{\nu^2 L_z+C_z\kappa^2}{\omega_{min}^4}}\right)\label{eq:laggrowth}
\end{equation}
in terms of $\omega_{min}^2$ defined in the main text.  Instability can thus identified once again from $\omega_{min}^2<0$, yielding an identical stability threshold as determined in the absence of a rotation lag.   However, now the condition $(C_z\kappa^2+\nu^2 L_z)>0$ must also be met.  This yields a condition on $k_z$ for instability (substituting in the expression for $L_z$ defined in eq. [\ref{eq:lz}]), i.e.

\begin{equation}
\rho_1\kappa^2k_z^2\left(\frac{-4\pi G\rho_0}{k^2+k_z^2}+\sigma_z^2\right)>-kk_z\frac{z}{R_\nu}\rho_1\nu^2\left(\frac{-4\pi G\rho_0}{k^2+k_z^2}+\sigma_z^2\right)
\end{equation}
or
\begin{equation}
k_z>k z\frac{\nu^2}{\kappa^2}\frac{1}{R_{0}}\label{eq:kzconditionlag}
\end{equation}
where $R_{\nu}$ is approximated as $-R_0$ assuming that background density falls off approximately exponentially with scale length $R_0$.  

The above condition is most easily met precisely at the mid-plane ($z=0$) with vanishing rotational lag.  Elsewhere, it adds a negligible constraint when the background density distribution is a slowly decreasing function of $R$ such that $kR_0>>1$, as it is indeed assumed when invoking the WKB approximation in the radial direction.

At a small distance $z$ above the above the mid-plane, eq. (\ref{eq:kzconditionlag}) can be used to place a condition on $k$, given an accessory requirement $k_zh<<1$, i.e.

\begin{equation}
k<\frac{1}{h}\frac{R_0}{z}\frac{\kappa^2}{\nu^2}.
\end{equation}
A rotational lag thus places a height-dependent minimum on the wavelength of the radial perturbations that can lead to instability, adding together with the condition on $k$ determined from identifying when $\omega_{min}^2<0$.  The latter is the stronger constraint assuming $1/R_{\nu}$ is indeed small.  

Notice that, according to eq. (\ref{eq:laggrowth}), the growth rates of perturbations can be slowed in the presence of a rotational lag, depending on vertical stability.   Wherever $C_z>0$ and the vertical direction is unstable, $L_z<0$.  As the lag term $\vert L_z\vert$ increases, $\omega^2$ decreases until the point $L_z>-C_z\kappa^2/\nu^2$ and real solutions are no longer permitted.  
For very large $d V_c/dz$, our adopted 3D WKB perturbations would no longer satisfy the equations of motion. Indeed, for large enough $\vert L_z\vert$, the perturbed radial velocity in eq. (\ref{eq:radvelocity}) is dominated less by self-gravity (and pressure) and more by the outward motion associated with moving up in the weakening potential.   

\subsection{In the Overall Disk}
The $z$-dependence of the lag term $L_z$ in the previous section gives a non-negligible rotational lag very little influence on the overall stability of the disk, since 

\begin{equation}
\int_{-\infty}^{\infty}\nu^2L_z dz=0
\end{equation}
and 
\begin{equation}
\int_{-h_1}^{h_1}\nu^2L_z dz\approx 0 \hspace*{.5cm} 
\end{equation}
for either $h_1>>h$ or for $h_1<<h$.   
Thus, the lag term drops from the 2D dispersion relation for infinite wave and non-wave perturbations and for all finite WKB perturbations, leaving the stability conditions exactly as determined in $\S$ \ref{sec:2dstability}.  

Indeed, the variation in $\kappa$ with height above the mid-plane implied by the presence of a rotational lag introduces negligible change in the overall stability threshold in these cases.  
As an illustration, take $\kappa=\sqrt{2}\Omega$ in the flat part of the rotation curve, which implies

\begin{eqnarray}
\frac{d\kappa^2}{dz}&=&2\sqrt{2}\frac{\Omega}{R}\frac{dV_c}{dR}\\
&\approx&\frac{\sqrt{2}}{R}\frac{d}{dR}\nu^2z
\end{eqnarray}
from which it can be estimated that

\begin{eqnarray}
\kappa^2(z)&=&\kappa^2(z=0)+\int \frac{d\kappa^2}{dz} dz\\
&=&\kappa^2(z=0)-\frac{\sqrt{2}}{R}\frac{z^2}{2}\frac{\nu^2}{R_0}
\end{eqnarray}
with $R_0$ as used in the previous section.  
In this case, the perturbation-weighted $\kappa^2$ that would appear in the 2D dispersion relation is

\begin{eqnarray}
\bar{\kappa}^2&=&\frac{\int_{-\infty}^{\infty}\kappa^2(z)\rho_1 dz}{\int_{-\infty}^{\infty}\rho_1 dz}\\
&=&\kappa^2(z=0)-\frac{h^2\nu^2}{\sqrt{2}RR_0}\\
&=&\kappa^2(z=0)-\frac{\sigma^2}{\sqrt{2}RR_0}
\end{eqnarray}
The first term by far dominates in the gas disks of nearby galaxies since $V_c>>\sigma$.  Even in extremely puffy disks with $V_c/\sigma$=2, $\bar{\kappa}^2$ easily remains within a factor of 2 of $\kappa^2(z=0)$ within $4R_0$.  (Note, though, that such puffy-disk scenarios are unlikely a good match for the weakly-self-gravitating assumption adopted for this approximation.)   

\section{Stability of (Less) Tightly-wound Non-axisymmetric Perturbations}\label{sec:nonaxisym}
In this section we appeal to linear theory to examine the stability of disks to non-axisymmetric ($m\neq$0) perturbations as considered by \cite{GLBb, JT66, laubertin, bertinplus, grivgedalin}.  Introducing the azimuthal forces associated with these perturbations involves a weakening of the requirement that $k R>>1$ normally adopted with the WKB approximation.  The result is a picture of the destabilizing influence of azimuthal forces that lead to growth, in the manner ultimately described by swing amplification \citep{GLBb, JT66,toomre81}.  A basic diagnostic of this behavior is an increase in the $Q_T$ threshold for stability, as shown in a  number of studies.  In order to compare this change to the increase from $Q_T$=1 to $Q_T=2/(\alpha f_g^{1/2})$ predicted endemic to the mid-plane ($\S\S$ \ref{sec:3dmidplane} and \ref{sec:smallh1}) the calculation of $Q_T$ for $m>0$ is reproduced here, adopting the 3D perturbations and framework described in the main text.  

The first steps involve substituting the full expressions for $v_{r,1}$ and $v_{\theta,1}$ in eqs. (\ref{eq:radvelocity}) and (\ref{eq:phivelocity}) into the continuity equation (eq.[\ref{eq:fullcontinuity}]).  Here, the perturbed pressure term is written in terms of the enthalpy $\eta_1$, i.e. setting $\eta_1=\sigma^2\rho_1/\rho_0$.  

\begin{eqnarray}
0&=&-(\omega-m\Omega)\frac{\rho_1}{\rho_0}\nonumber\\
&-&\frac{(\Phi_1+\eta_1)}{\Delta}\Big[(\omega-m\Omega)k^2+\frac{2m\Omega}{R^2}\left(1+R\frac{\partial\ln\Sigma_0}{\partial R}\right)\nonumber\\
&+&2m\frac{d\Omega}{dR}+i\frac{k}{R}(\omega-m\Omega)\left(1+R\frac{\partial\ln\Sigma_0}{\partial R}\right)-i\frac{k}{R}2m\Omega\Big]\nonumber\\
&-&\frac{(\Phi_1+\eta_1)}{\Delta}\Big[\frac{m^2}{R^2}(\omega-m\Omega)-2iBk^2\Big]\nonumber\\
&+&\frac{C_z}{\rho_0(-\omega+m\Omega)}\label{eq:2Ddispwithm}
\end{eqnarray}
where the second and third terms originate with the radial and azimuthal components of the velocity, respectively.  The rotational lag terms have been neglected (see Appendix \ref{sec:rotationallagappendix}).  

From this point, \cite{laubertin} argue that that the out-of-phase terms arising with the in-plane imaginary parts of eq. (\ref{eq:2Ddispwithm}) are not important for stability and growth and can be neglected.  The continuity equation thus becomes

\begin{eqnarray}
0=\frac{\rho_1}{\rho_0}\Delta&+&\left(k^2+\frac{m^2}{R^2}\right)(\Phi_1+h_1)\nonumber\\
&+&\left[2m\frac{\Omega}{R(\omega-m\Omega)}\left(\frac{d\ln\Omega}{dR}-\frac{d\ln\Sigma_0}{d R}\right)\right](\Phi_1+\eta_1)\nonumber\\
&+&\frac{\Delta C_z}{\rho_0(-\omega+m\Omega)^2}\label{eq:full3dnonaxi}
\end{eqnarray}

Another simplification involves continuing to require that the characteristic scale of variation in the perturbation's amplitude is small compared to $1/k$ and tied to the unperturbed disk. Following \cite{morozov}, then, we assume $kL>>1$ where $L=\textrm{min}(\vert d\ln\Omega/d R\vert^{-1},\vert d\ln\Sigma_0/d R\vert^{-1})$, such that the term in square brackets can be neglected.  

Before examining the 3D dispersion relation at the mid-plane in $\S$ \ref{sec:3dnonaxi}, for reference 2D dispersion relation derived by adopting a delta function perturbation $\rho_1=\Sigma_1\delta(z)$ is first presented below.  In this case it is typical to let $\Phi_1=(2\pi G\Sigma_1/k) e^{-kz}$ (see BT), neglecting disk thickness.  

\subsection{2D Stability using Delta Function Perturbations}\label{sec:2ddelta}
In the absence of perturbation that entails explicit vertical motion, integration of the continuity equation yields the 2D dispersion relation 

\begin{equation}
(\omega-m\Omega)^2=\kappa^2+\left(-\frac{2\pi G\Sigma_0}{k}+\sigma_r^2\right)\left(k^2+\frac{m^2}{R^2}\right).\label{eq:nonaxidispersion}
\end{equation}

Now with the requirement $(\omega-m\Omega)^2<0$ sufficient for identifying the condition $\omega^2<0$ for growth, the conditions on $k$ for instability can be identified from 

\begin{equation}
\kappa^2-2\pi G\Sigma_0 k\left(1+\frac{m^2}{k^2R^2}\right)+\sigma_r^2 k^2\left(1+\frac{m^2}{k^2R^2}\right)<0.\label{eq:nonaxigrowth}
\end{equation}

At this stage, it is typical to effectively assume that the pitch angle $i_p$ of the perturbation is unvarying, such that the quantity $m^2/(kR)^2=\tan^2{i_p}$ is roughly constant.  Thus eq. (\ref{eq:nonaxigrowth}) can be easily solved for $k$, i.e.  

\begin{equation}
k<\frac{\pi G\Sigma_0}{\sigma_r^2}\left[1\pm\left(1-\frac{Q_T^2}{\left(1+\tan^2(i_p)\right)}\right)^{1/2}\right]
\end{equation}
yielding the stability criterion
\begin{equation}
Q_T>\left(1+\tan^2(i_p)\right)^{1/2}.\label{eq:Qnonaxi}
\end{equation}

This is identical to the $Q_T$ threshold derived by \cite{grivgedalin} to lowest order in $m^2/(kR)^2$ in the case of a flat rotation curve. (Note that eq. (\ref{eq:nonaxigrowth}) in the limit $kR<<m$ to lowest order in $k$ implies that the disk is always unstable and there is no $Q_T$ threshold for exceptionally loose perturbations.) 

Eq. (\ref{eq:Qnonaxi}) is also equivalent to the change in $Q_T$ threshold found in the presence of non-axisymmetric structure by \cite{laubertin} \citep[and][]{bertinplus} when substituting the value of $k$ associated with the most unstable mode, i.e $k=2\pi G\Sigma_0/\sigma_r^2=2\kappa/(Q_T\sigma_r)$. In this case, stability requires 

\begin{equation}
Q_T>\left(1+\frac{m^2\sigma_r^2 }{4\kappa^2R^2}\right)^{1/2}
\end{equation}
to lowest order in $1/(\kappa R)$ or  
\begin{equation}
Q_T>\left(1+\frac{m^2\sigma_r^2 }{8V_c^2}\right)^{1/2}\label{eq:QnonaxiLB}
\end{equation}
when the rotation curve is flat and $\kappa=\sqrt{2}V_c/R$.  

The $Q_T$ threshold is thus generally raised for tightly wound non-axisymmetric structures.  However, the increase estimated here is negligible for most scenarios, and the $Q_T$=1 threshold mostly remains accurate \citep{BT}. 
In the stellar disks of nearby galaxies with $\sigma_r/V_c\sim0.2$ or lower, the change to the $Q_T$ threshold is only appreciable for the very loosest perturbations ($Q_T$$\lesssim$1.2 for all $m<10$). In gas disks with even lower $\sigma_r/V_c\lesssim0.1$, the stability threshold is raised to $Q_T\sim1.5$ only for $m\gtrsim 30$ (although eq. (\ref{eq:Qnonaxi}) looses its accuracy for such loose perturbations.) 

\subsection{3D (In)stability at the Mid-plane}\label{sec:3dnonaxi}
Now consider a 3D perturbation that is WKB-like near the mid-plane with non-axisymmetry in the plane such that 

\begin{equation}
\Phi_1=\frac{4\pi G\rho_1}{k^2+\frac{m^2}{R^2}+k_z^2}.
\end{equation}
(from Poisson's equation). Now substituting in eq. (\ref{eq:czmidplane}) into eq. (\ref{eq:full3dnonaxi}), the 3D dispersion relation is once quadratic in $\omega^2$, but now with the addition of the in-plane non-axisymmetric terms. Following the arguments in $\S$~\ref{sec:3dmidplane}, the condition for instability in this case becomes

\begin{eqnarray}
0&>&\kappa^2+\left(-\frac{4\pi G\rho_0}{k^2+\frac{m^2}{R^2}+k_z^2}+\sigma^2\right)\left(k^2+\frac{m^2}{R^2}\right)\nonumber\\
&+&\left(-\frac{4\pi G\rho_0}{k^2+\frac{m^2}{R^2}+k_z^2}+\sigma^2\right)k_z^2
\end{eqnarray} 
or
\begin{equation}
0>\kappa^2-4\pi G\rho_0+\sigma_r^2k^2+\sigma_r^2\frac{m^2}{R^2}+\sigma_z^2k_z^2.
\end{equation}

Once again adopting the assumption of a fixed pitch angle, instabillity is possible as long as

\begin{equation}
k^2<\frac{k_J^2(1-Q_M-k_z^2h^2)}{\left(1+\tan^2{(i_p)}\right)}
\end{equation}
provided that $Q_M<1$ in the limit $k_zh<<1$.  

Dropping the fixed $i_p$ assumption in practice yields a similar $Q_M$ threshold. Instability would proceed where

\begin{equation}
k^2<k_J^2(1-Q_M-k_z^2h^2)-\frac{m^2}{R^2}
\end{equation}
suggesting the stability threshold $Q_M=1-m^2h^2/R^2$ (in the limit $k_zh<<1$), which is equivalent to $Q_M\approx 1$ for  thin gas disks.  The introduction of non-axisymmetry is thus of negligible impact on the mid-plane stability threshold.

\end{document}